\documentclass[10pt,letterpaper]{article}
\usepackage[top=0.85in,left=1in,footskip=0.75in,marginparwidth=2in]{geometry}

\usepackage[utf8]{inputenc}
\usepackage{amsmath,amssymb,bbm}
\usepackage[shortlabels]{enumitem}

\usepackage{float, graphicx}
\usepackage[ruled,vlined,linesnumbered]{algorithm2e}

\usepackage{color, xcolor}

\usepackage[aboveskip=1pt,singlelinecheck=off,font=small,labelfont=bf, labelsep=period]{caption}
\usepackage{subcaption}

\usepackage[backend=biber, style=ieee, sorting=ynt, maxcitenames=3,style=alphabetic]{biblatex}
\DeclareDelimFormat{nameyeardelim}{\addcomma\space} 
\usepackage{csquotes}
\SetCiteCommand{\autocite} 
\usepackage{hyperref} 

\usepackage{cleveref}
\Crefname{algocf}{Algorithm}{Algorithms} 
\Crefname{paragraph}{Paragraph}{Paragraphs}

\raggedright
\usepackage{lastpage,fancyhdr}
\newcommand \RR {\mathbb{R}}

\DeclareMathOperator*{\argmin}{arg\,min}

\DeclareMathOperator*{\ES}{ES} 

\newcommand{\code}[1]{\verb|#1|}

\usepackage{awesomebox}
\newcommand{\conclulogo}[1]{\awesomebox{0pt}{\faMarker}{black}{\textbf{Conclusion:} #1}}
\newcommand{\infologo}[1]{\awesomebox{0pt}{\faInfoCircle}{black}{\textbf{Information:} #1}}
\newcommand{\warninglogo}[1]{\awesomebox{0pt}{\faExclamationTriangle}{black}{\textbf{Warning:} #1}}

\usepackage[most]{tcolorbox}
\newtcolorbox{blackbox}[1]{colback=white, colframe=black, 
coltext=black, boxsep=1.5pt, arc=4pt, before=\centering, title=#1}

\newtcolorbox{info}[1]{colback=green!15!white,  coltitle =white, colframe=green,
 coltext=black, boxsep=1.5pt, arc=5pt, before=\centering, 
 title={\infologo{#1}}, drop shadow, boxrule=1mm}
\newtcolorbox{warning}[1]{colback=red!15!white, colframe=red, 
coltext=black, boxsep=1.5pt, arc=5pt, before=\centering,
title={\warninglogo{#1}}, drop shadow, boxrule=1mm}
\newtcolorbox{conclusion}[1]{colback=orange!15!white, colframe=orange, coltext=black,
boxsep=1.5pt, arc=5pt, before=\centering, title={\conclulogo{#1}},
 drop shadow, boxrule=1mm}

\usepackage{multicol} 
 \tcbuselibrary{theorems} 
\tcbset{ 
defstyle/.style={fonttitle=\bfseries\upshape, fontupper=\slshape,
arc=2pt, colback=blue!5!white,colframe=blue!75!black},
theostyle/.style={fonttitle=\bfseries\upshape, fontupper=\slshape,
colback=red!10!white,colframe=red!75!black, arc=2pt, every float=\centering},
propstyle/.style = {fonttitle=\bfseries\upshape, fontupper=\slshape,
 arc=2pt, colback=green!5!white,colframe=green!75!black},
proofstyle/.style = {colback=white,
  colframe=black,  coltext=black, arc=2pt}
}

\newtcbtheorem[number within=section,Crefname={definition}{definitions}]%
{Definition}{Definition}{defstyle}{def}
\newtcbtheorem[use counter from=Definition,Crefname={theorem}{theorems}]%
{Theorem}{Theorem}{theostyle}{th}
\newtcbtheorem[use counter from=Theorem,Crefname={Proof}{Proofs}]%
{Proof}{Proof}{proofstyle}{proof}
\newtcbtheorem[use counter from=Definition,Crefname={corollary}{corollaries}]%
{Corollary}{Corollary}{theostyle}{cr}
\newtcbtheorem[number within=section,Crefname={property}{properties}]%
{Property}{Property}{propstyle}{pr}

\begin{document}
\vspace*{0.35in}

\begin{flushleft}
{\Large
\textbf
\newline{An updated State-of-the-Art Overview of transcriptomic Deconvolution Methods}
}
\newline
\\
Bastien Chassagnol\textsuperscript{1,2,*},
Grégory Nuel\textsuperscript{2},
Etienne Becht\textsuperscript{1}
\\
\bigskip
\bf{1} Institut De Recherches Internationales Servier (IRIS), FRANCE
\\
\bf{2} LPSM (Laboratoire de Probabilités, Statistiques et Modélisation), Sorbonne Université, 4, place Jussieu, 75252 PARIS, FRANCE
\\
\bigskip
* bastien\_chassagnol@laposte.net

\end{flushleft}

\section*{Abstract}
Although bulk transcriptomic analyses have significantly contributed to an enhanced comprehension of multifaceted diseases, their exploration capacity is impeded by the heterogeneous compositions of biological samples. Indeed, by averaging expression of multiple cell types, RNA-Seq analysis is oblivious to variations in cellular changes, hindering the identification of the internal constituents of tissues, involved in disease progression. On the other hand, single-cell techniques are still time, manpower and resource-consuming analyses.

To address the intrinsic limitations of both bulk and single-cell methodologies, computational deconvolution techniques have been developed to estimate the frequencies of cell subtypes within complex tissues. These methods are especially valuable for dissecting intricate tissue niches, with a particular focus on tumour microenvironments (TME).

In this paper, we offer a comprehensive overview of deconvolution techniques, classifying them based on their methodological characteristics, the type of prior knowledge required for the algorithm, and the statistical constraints they address. Within each category identified, we delve into the theoretical aspects for implementing the underlying method, while providing an in-depth discussion of their main advantages and disadvantages in supplementary materials.

Notably, we emphasise the advantages of cutting-edge deconvolution tools based on probabilistic models, as they offer robust statistical frameworks that closely align with biological realities. We anticipate that this review will provide valuable guidelines for computational bioinformaticians in order to select the appropriate method in alignment with their statistical and biological objectives.

We ultimately end this review by discussing open challenges that must be addressed to accurately quantify closely related cell types from RNA sequencing data, and the complementary role of single-cell RNA-Seq to that purpose.

\clearpage

\section{Introduction}  

\subsection{Main sources of transcriptomic variability}

The transcriptome refers to the complete set of RNA transcripts, expressed within a biological sample. By providing a snapshot of gene expression patterns, studying its variations across phenotypical conditions provide valuable insights into the regulatory mechanisms of gene expression that underlie disease progression and individual responses to treatments.

\begin{figure}[H]
         \centering
         \includegraphics[width=0.7\textwidth]{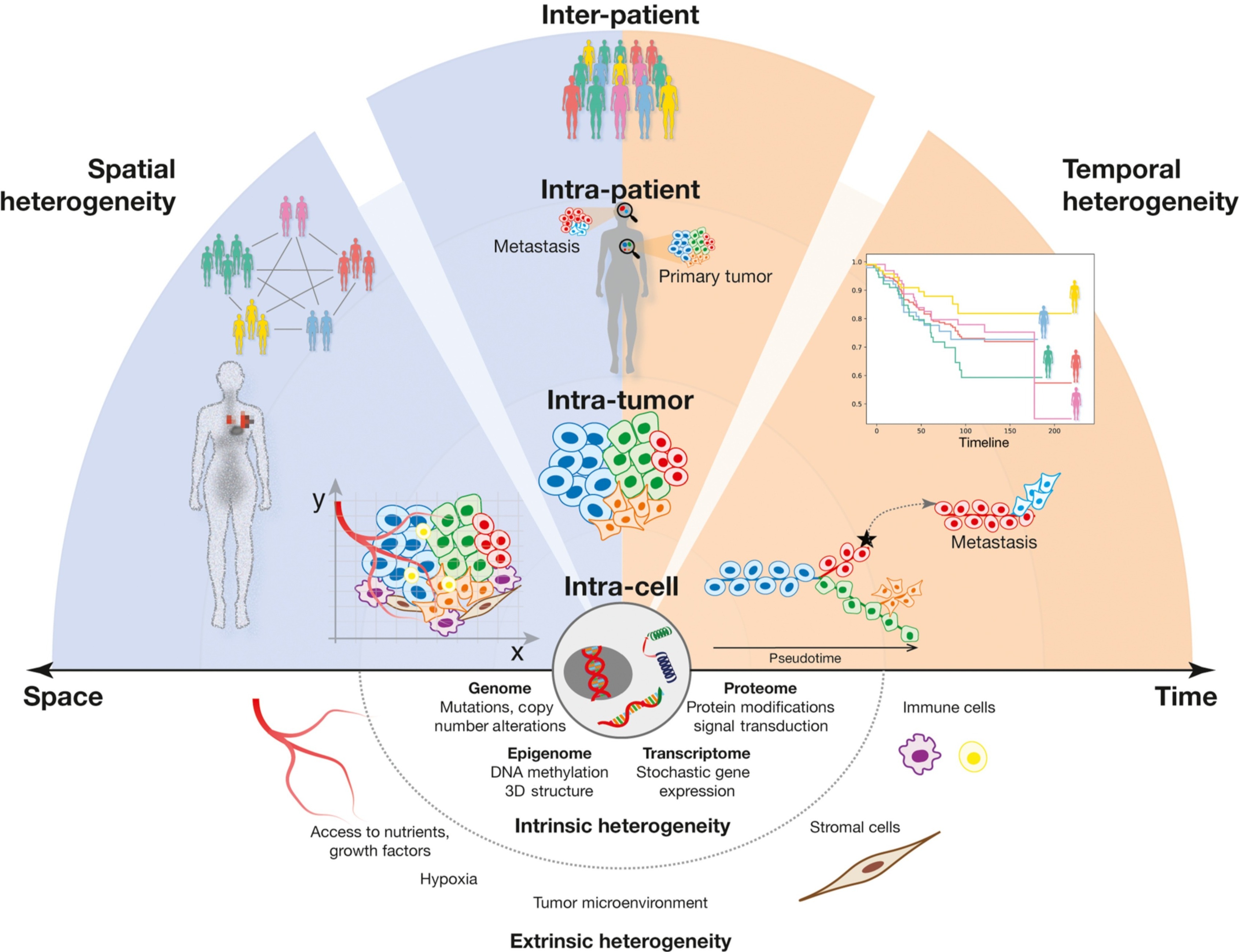}   
         \caption{\textbf{Main sources of transcriptomic variability, illustrated by the the intricacy of tumoral environments.} The diversity of molecular profiles proceeds from a combination of intrinsic and extrinsic factors. \textit{Intrinsic factors} encompass stochastic genetic, transcriptional, and proteomic mechanisms, while \textit{extrinsic factors} include interactions between the resident cell populations and the surrounding microenvironment. The interconnection between these factors requires a systematic and multi-layered approach to comprehensively understand the intricacy of such biological environments. Figure reproduced from \autocite[Fig. 1]{kashyap_etal22}} 
          \label{fig:deconv-heterogeneity}
\end{figure}

The main biological sources of transcriptomic expression, between individuals and within tissues, proceed from three main biological factors, summarised in \Cref{fig:deconv-heterogeneity}: the global environmental and topic condition of the sample, encompassing disease state and tissue location; the genotype condition, involving single-nuclear polymorphisms, haplotypes, and comparable genetic aspects; and the cellular composition. Changes of cell composition are notably driven by intertwined physiological processes activating \emph{cell motility} and \emph{cell differentiation} mechanisms (\autocite{shen-orr_gaujoux13}). In addition, the pertinent biological signal is often entangled with extraneous technical noise, requiring specific corrections in subsequent downstream analyses.

In addition, intrinsic heterogeneity is also present at the cell population level itself, arising from the presence of unspecified and infrequent population subtypes, coexistence of different developmental \textit{cell states} or asynchronous biological processes (such as the cell cycle or circadian rhythm). Lastly, the kinetics of transcriptome regulation is inherently stochastic \autocite{buettner_etal15} (see \Cref{fig:deconv-heterogeneity}).

While the analysis of the transcriptome through bulk RNA-Seq reveals meaningful co-expression patterns, by averaging measurements over several cell populations, it tends to ignore the intrinsic heterogeneity and complexity inherent to biological samples. Accordingly, bulk RNA-based methods are usually not able to determine whether significant changes in gene expression stem from a change of cell composition, from phenotype-induced variations or a combination of these factors (\autocite{kuhn_etal12}).

\begin{figure}[H]
     \centering
         \includegraphics[width=0.8\textwidth]{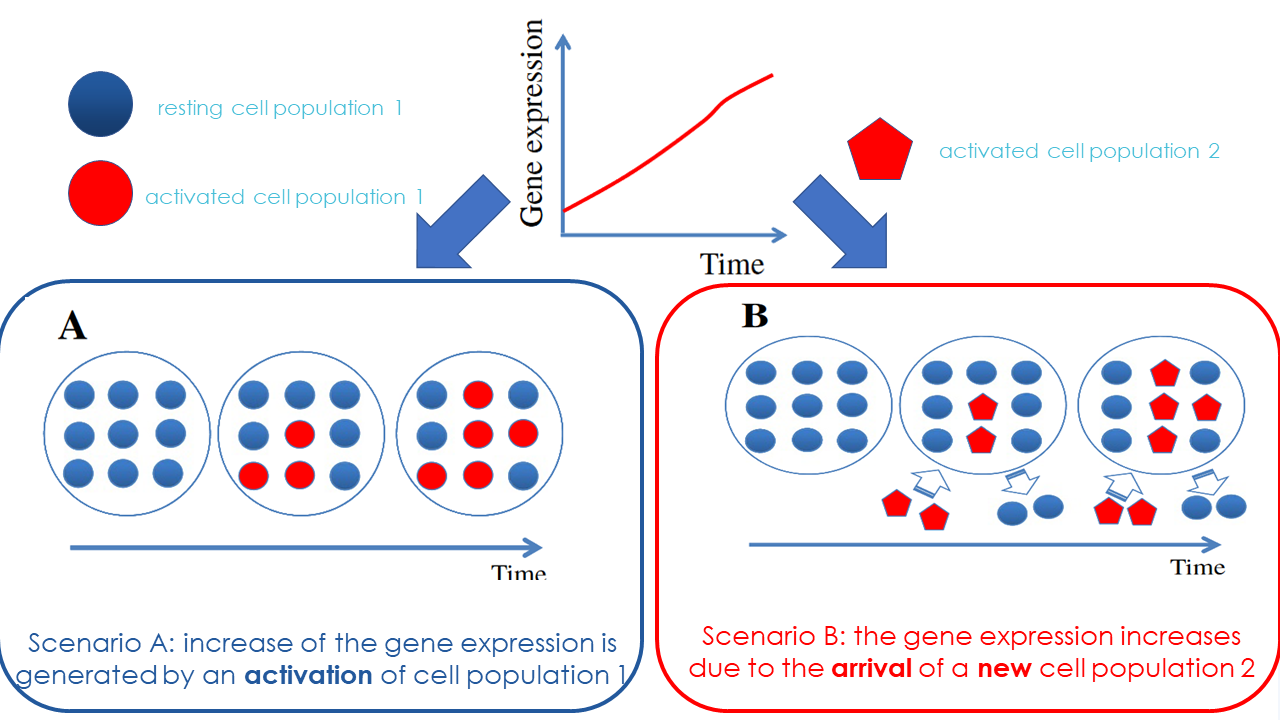}
         \label{fig:deconv-role}
         \caption{\textbf{Changes in cell composition impact the transcriptomic expression.} Here, at least two distinct biological mechanisms can likely explain the increased expression of transcriptomic activity observed for a given marker gene. In the scenario (A), the cell composition is unchanged, but previously inactivated cells are stimulated and released the TF in the biological medium. In scenario (B), there is a change of cell composition, with the infiltration of a second cell type in the sample. Reproduced from \autocite[Fig. 1]{shoemaker_etal12}.} 
\end{figure}

Hence, failure to account for changes of the cell composition is likely to result in a loss of \emph{specificity} (genes mistakenly identified as differentially expressed, while they only reflect an increase in the cell population naturally producing them) and \emph{sensibility} (genes expressed by minor cell populations are amenable being masked by highly variable expression from dominant cell populations), as simply illustrated in \Cref{fig:deconv-role}. 
Overall, the intrinsic heterogeneity of complex tissues, above all tumoral ones, reduces the robustness and reproducibility of downstream analyses, notably differential gene expression analysis or clustering of co-expression networks \footnote{\autocite{whitney_etal03} notably exhibits that most of the variability of gene expression in whole blood samples proceeds from relative changes of the composition in neutrophils, the most abundant immune cell type.}.

Various computational methodologies have emerged in recent years to estimate automatically cell type proportions in biological samples from bulk transcriptomic profiles, alleviating the high costs of single-cell RNA-Seq technologies or enabling the exploitation of archived patient datasets whose original material is not anymore available \autocite{avilacobos_etal18}. Furthermore, by requiring prior isolation of cell populations single-cell technologies hinder the analysis of interactions occurring between them.

In contrast to bulk RNA-Seq and single cell methodologies, computational techniques can simultaneously capture systemic and cell-specific information, respectively (\autocite{shen-orr_gaujoux13}). Accordingly, by dissecting the intricacy of tissues, they reveal a strong potential to identify causal drivers and provide insights on regulation mechanisms. 

\subsection{Overview of numerical deconvolution methods}

\begin{figure}[H]
         \centering
         \includegraphics[width=0.6\textwidth]{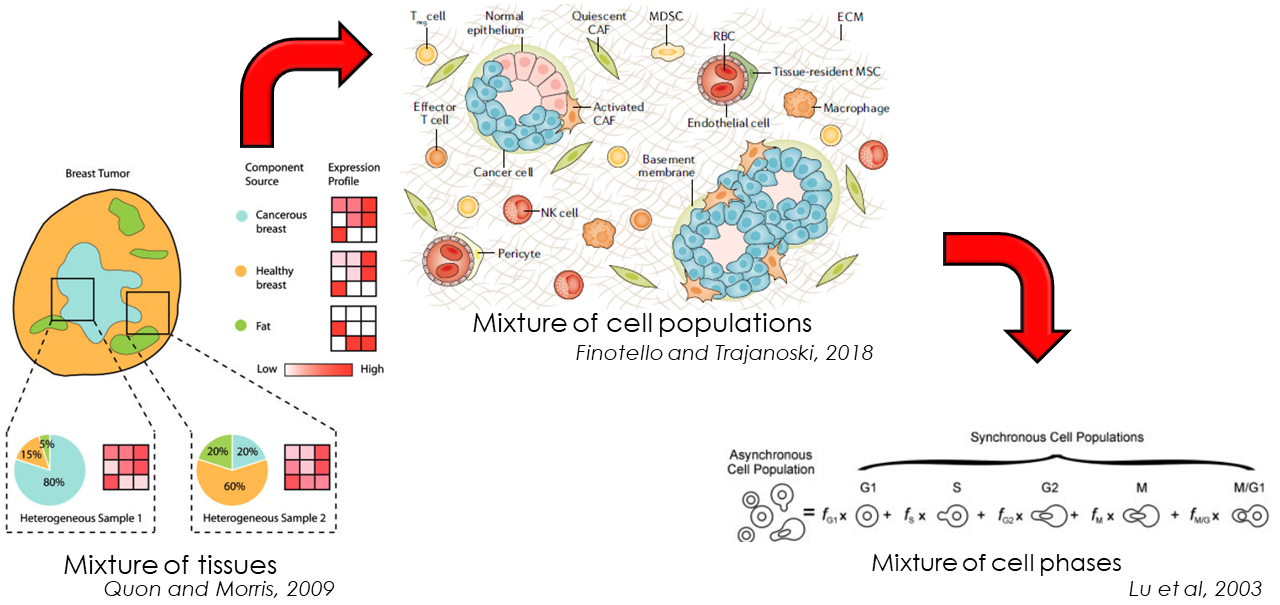}
         \caption{We detail some common applications of deconvolution methods, ordered by tier of resolution, from the least detailed resolution: \textit{tissue} level (\autocite[Fig .1]{quon_morris09}), to the most detailed one, \textit{cell cycles} (\autocite[Fig .1]{lu_etal03}), through the \textit{cell population} strata (\autocite[Fig .1]{finotello_etal19}).} 
         \label{fig:deconv-applications}         
\end{figure}

\emph{Deconvolution} generally speaking names the process that consists
in retrieving from a mixture its individual sub-components, popularised as the \enquote {cocktail party problem} \autocite{cherry53}. In a biological sample
(whole blood, tissue, \ldots), this consists generally in retrieving the
distinct cell populations (immune, stromal\ldots) composing it, but it
can be directly extended to identify the different sources of the RNA
production (for instance, many studies investigate on estimating a
tumour purity score returning the proportion of malignant cells in \autocite{yoshihara_etal13}) or, at higher resolution, identify the cycle stages within a cell population (see \Cref{fig:deconv-applications}).

Traditionally, deconvolution models assume that the total bulk expression is linearly related to the individual cell profiles. Precisely, they posit that the global expression can reconstructed by summing the distinct contributions of every cellular population weighed by their respective abundance within the sample (see \Cref{eq:linear-deconvolution} and graphical illustration in \Cref{subfig:linear-deconvolution}):

\begin{equation}
\begin{split}
	\boldsymbol{y}_i  &= \boldsymbol{X} \times \boldsymbol{p}_i \quad \text{matricial form}\\
	 y_{gi}  &=\sum_{j=1}^J x_{gj} \times p_j \quad \text{algebraic form}
\end{split}
\label{eq:linear-deconvolution}
\end{equation}, with the following notations:
\begin{itemize}
    \item $(\boldsymbol{y}=(y_{gi}) \in \mathbb{R}_+^{G\times N}$ is the global bulk transcriptomic expression, measured in $N$ individuals.
    \item  $\boldsymbol{X}=(x_{gj}) \in \mathcal{M}_{\RR^{G\times J}}$ the signature matrix of the mean expression of $G$ genes in $J$ purified cell populations.
    \item $\boldsymbol{p}=(p_{ji})\in ]0, 1[^{J \times N}$ the unknown relative proportions of cell populations in $N$ samples
\end{itemize} 

Overall, the system includes $G$ linear equations with $J$ unknowns (the cellular proportions). In addition, most deconvolution problems explicitly integrate the \textit{compositional} nature of cell ratios, enforcing in the estimation process the \textit{unit-simplex constraint} (\Cref{eq:simplex-constraint}):

\begin{equation}
\begin{cases}
\sum_{j=1}^J p_{ji}=1\\
\forall j \in \widetilde{J} \quad p_{ji}\ge 0
\end{cases}
\label{eq:simplex-constraint}
\end{equation} 

Implicitly, \Cref{eq:simplex-constraint} implies that no other, unknown cell population could contribute to the measured bulk mixture. The main classes of deconvolution methods, defined on the basis of their biological objectives, are summarised in \Cref{subfig:deconvolution-types-classes}, ranging from the approaches requiring the most information to the most unsupervised approaches:

\begin{figure}
     \begin{subfigure}[H]{0.9\textwidth}
     \centering
         \includegraphics[width=0.7\textwidth]{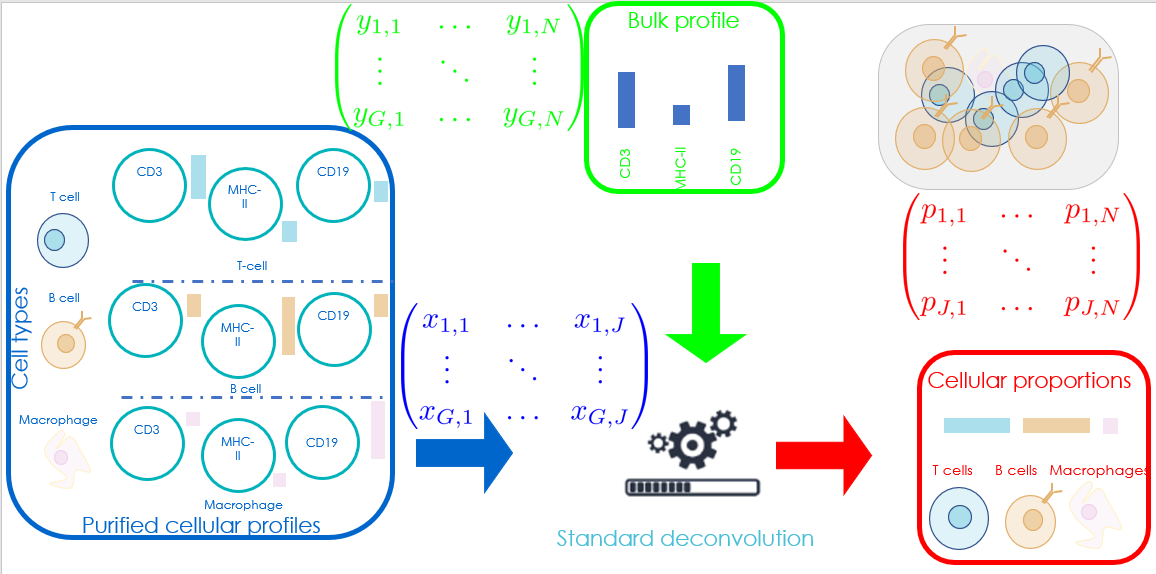}
         \label{subfig:linear-deconvolution}
         \caption{Graphical abstract, illustrating the fundamental linear assumption of bulk mixture construction underlying the cellular deconvolution framework.} 
     \end{subfigure}
     \vfill
     \begin{subfigure}[H]{0.9\textwidth}
     \centering
         \includegraphics[width=0.5\textwidth]{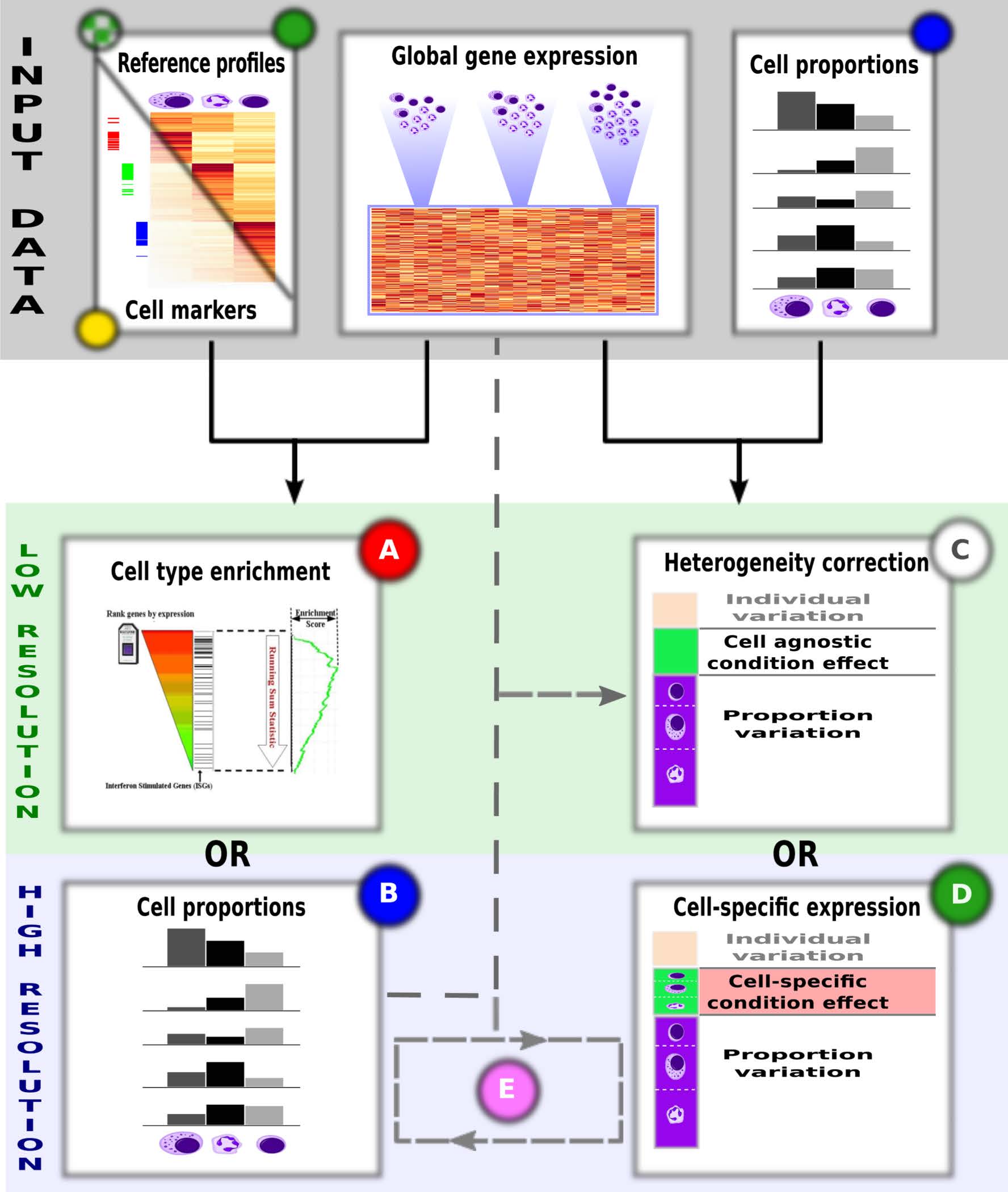}         
         \caption{The deconvolution methods are classified according to their input data requirements as well as the output type and resolution they provide. Supervised, alternatively named partial methods, methods utilise markers, signatures, or cytometry proportions, to achieve cell detection (A), estimating cell proportions (B), correcting heterogeneity (C), or estimating cell type-specific expression profiles (D), ranked from the simplest to the most challenging task.On the other hand, complete deconvolution methods (E) simultaneously estimate cellular proportions and purified expression profiles. Reproduced from \autocite[Fig. 3]{shen-orr_gaujoux13}.} 
         \label{subfig:deconvolution-types-classes}
     \end{subfigure}
\end{figure}

In the following \Cref{sec:reference-based}, we focus on \emph{partial deconvolution} methods, that require individual cellular expression profiles to infer cell composition \autocite{stuart_etal04}. Besides, in the remainder of this paper, we posit, as most deconvolution algorithms, that the samples are uncorrelated with each other (independence assumption), allowing simultaneous and parallel cell ratio estimations. While this assumption reduces computational complexity, \autocite{efron09} demonstrates cross-correlation across samples in real-world transcriptomic profiles.

\section{Reference-based Approaches: Deciphering Cell Mixture through Expression Signatures}
\label{sec:reference-based}

\subsection{Regression-based approaches}
\label{par:linear-modelling-cellular-ratios}

The system of linear equations, given in \Cref{eq:linear-deconvolution} rarely holds in practice, due to technical noise or unaccounted environmental variations. Most deconvolution algorithms model explicitly the error with a residual unobserved term, added to each individual transcriptomic measure,  $\epsilon_{g}$.

Subsequently, the usual approach is to retrieve the ordinary least squares (OLS) estimate which minimise the sum of squares (SSE) between predicted values fitted by the linear model: \(\hat{\boldsymbol{y}} = \boldsymbol{X}\hat{p}\) and the actually observed and measured values: \(\boldsymbol{y}\): 

\begin{equation}
 \boldsymbol{\hat{p}}^{\text{OLS}} \equiv \argmin_{\boldsymbol{p}} ||\hat{\boldsymbol{y}} - \boldsymbol{y}||^2 = \argmin_{\boldsymbol{p}}  ||\boldsymbol{X}\boldsymbol{p} - \boldsymbol{y}||^2 = \sum_{g=1}^G \left( y_{g} - \sum_{j=1}^J  x_{gj} p_{j}\right)^2
 \label{eq:OLS-task}
\end{equation}

with $\hat{\boldsymbol{p}}$ the unknown \textit{coefficients} to estimate, \(\boldsymbol{y}\) known as the \textit{predicted, response variable} in a linear regression context and \(\boldsymbol{X}\) the \emph{design matrix}, storing the $J$ purified profiles. Note that the \enquote{Rouché-Capelli} theorem states that the uniqueness of a solution to \Cref{eq:OLS-task} requires that the number of genes is at least equal to the number of cell ratios to estimate. The OLS estimator,\(\hat{p}_{\text{OLS}}\) is explicitly given by the \emph{Normal equations} (see Theorem A.1):

Interestingly, if we consider a generative approach, in which the error term is described by a white-\emph{Gaussian} process (homoscedastic, null-centred), the \emph{Gaussian-Markov} theorem (see Theorem A.2) states that the OLS estimate is unique and equal to the Maximum Likelihood Estimate (see Proof A.3).

\bigskip
Linear modelling, whose cellular ratios are the ones given by the Normal Equations (Theorem A.1), has first been used as such in \autocite{abbas_etal09} paper, using the \href{https://rdrr.io/r/stats/lsfit.html}{lsfit} function.  The same method is used in \autocite{li_etal16a},to identify subgroups of melanomas characterised by varying levels of TCD8 subsets and correlate them with prognostic factors. To avoid accounting for tumoral cells when asserting ratios of infiltrated cells, only genes both highly correlated to the
cell types of the sample and negatively correlated to the \emph{tumour purity}, defined as the ratio of \emph{aneuploid cells} exhibiting a non canonical number of chromosomes. 

However, assumption of homoscedasticity of the residuals makes standard linear approaches sensitive to
outliers, while they do not endorse explicitly the unit-simplex constraint (\Cref{eq:simplex-constraint}), requiring posterior normalisation of the coefficients.

\subsubsection{Weighted linear approaches}

The presence of an unknown cell population might be relaxed by including a constant intersection term $p_0$, adding in practice a column of ones in the design matrix. To account for potential heteroscedascity (variance of the errors depends on the gene value), weighted linear approaches allow users to add prior weights to modify the \textit{leverage} (contribution) of each gene to the computation of the OLS estimate. Considering \(\boldsymbol{W}\) the diagonal matrix of weights, the Weighted version of the Least Square estimate is given by \Cref{eq:weighted-ols-estimate}:

\begin{equation}
\hat{p}_{\text{wOLS}} = (\boldsymbol{X}^\top \boldsymbol{W}\boldsymbol{X})^{-1}\boldsymbol{X}^\top \boldsymbol{W}\boldsymbol{y}
 \label{eq:weighted-ols-estimate}
\end{equation}

EPIC \autocite{racle_etal17} combines this weighted approach with the addition of a column characterising the tumour profile in the signature matrix. \autocite{racle_etal17} notably provides two signatures of circulating and tumour-infiltrating immune cells, CAFs (cancer-associated fibroblasts) and epithelial cells, respectively designed for whole-blood and solid tumoral tissues, aggregating bulk and scRNA-Seq data.  

Instead, the quanTIseq \autocite{finotello_etal19} algorithm integrates an additional constant intersection term to quantify the contribution of the unknown tumoral content. In addition, to address the issue of cell \enquote{drop-outs} (cell populations, generally infrequent and/or exhibiting a strong correlation with other cell types, that are wrongly estimated as absent), a heuristic approach is employed whereby the final Tregs estimate is computed as the average of two Tregs measures, in the presence and absence of the TCD4+ subset in the design matrix. Tregs are indeed highly correlated with TCD4+ cell populations. 

In weighted linear approaches, individual gene contributions are usually provided by the user. Without prior knowledge, the usual approach is then to give less importance to genes exhibiting strong variability within a cell population. However, assigning appropriate weights to each gene typically necessitates either prior knowledge or strong assumptions about the dataset's distribution. We subsequently review in next \Cref{par:feature-selection-deconvolution} robust linear regression methods that compute the weights or trim outlying gene expression in a automated manner.

\subsubsection{Robust Linear Regression and SVR Approaches for Automated Selection of Transcriptomic Markers}
\label{par:feature-selection-deconvolution}

In the previously described approaches, the inclusion of all genes in the regression framework may yield biased estimates when the expression of some genes significantly differ, due to significant changes of sequencing protocol or phenotype condition between the bulk mixture and purified expression profiles. Unfortunately, outlying genes in least-square approaches have the strongest influence on the parameters estimation, in reason of the Euclidean metric used to evaluate the prediction error.

Several robust methods, making a compromise between \textit{efficiency} and \textit{robustness} of the estimate, have been proposed.  They are usually classified into \emph{M-estimates} (see Definition A.4), whereby an adaptive function is enforced on the residuals, giving less weights to those with strong leverage, and \emph{LTS estimates}, where a user-provided ratio of aberrant genes is automatically identified and trimmed (see Definition A.5).

With both methods, the weights assigned to each observation depend on the estimator which in turn depend on the weights. As a result, the robust estimator must be computed sequentially, these methods are accordingly referred to as Iteratively Reweighted Least Squares (IRLS) approaches. Uniform weights are usually assigned to each observation, subsequently, a standard least regression estimate is computed. Once the OLS obtained, each observation is reweighted, using the transformation induced by the \textit{influence function}, and which usually depends on its leverage on the regression framework. The subsequent IRLS estimates are then computed with those new weights, and the process continues until convergence \autocite{yohai87}).  

\bigskip
The RCR (Robust Computational Reconstitution) deconvolution algorithm, by \autocite{hoffmann_etal06}, notably couples the LAD (see Definition A.5) regression framework while adhering to the unit-simplex constraint (\cref{eq:simplex-constraint}).

A variant of the LTS (least trimmed squares) approach has been implemented by the FARDEEP algorithm \autocite{hao_etal19}. It has notably been modified to ensure convergence towards a final set of trimmed observations, in a linearly growing number of iterations. However, the algorithm is highly sensitive to the tuning parameter that controls the final number of observations trimmed during the regression. And while convergence and consistency of the algorithm is guaranteed, there's no theoretical guarantee that the final estimate returned is indeed optimal.

Overall, all the variants proposed in this section are proned to overfitting. Indeed, since these weights are derived from the model's performance, they are highly sensible to dataset-specific patterns, leading to potential inconsistent and poor results on newly observed datasets. In addition, they are less efficient than the standard OLS estimate in case the Gauss-Markov assumptions hold.  

\bigskip
Support-vector-regression are supervised machine learning algorithm featuring an alternative strategy to select genes. It turned out that in real-world experiences, they tend to exhibit increased robustness to noisy observations.
The first historical mention to SVR approach, termed \(\epsilon\)-SVR \autocite{cortes_vapnik95}, uses
a insensitive loss function, whose parameter $\epsilon$ is provided by the user to control the error rate tolerated on the outputs (see Definition A.6).

CIBERSORT (Cell Type Identification By Estimating Relative Samples Of RNA Transcripts), developed by
\autocite{newman_etal15}, utilises the the \(\nu-\)SVR (\autocite{cc_cj02}) variant. Instead of optimising the precision (error rate tolerance), the $\nu$ parameter controls the proportion of Support Vectors integrated in the regression framework (\autocite{scholkopf_etal00}) \footnote{\autocite{cc_cj02} demonstrates the equivalence between the two approaches: increasing the \(\nu\) hyper-parameter results in a smaller \(\epsilon\)-tube and a higher precision on the results. Asymptotically, determining the \(\nu\)-proportion of support vectors reaching a given precision \(\hat{\epsilon}\), is even equal to the output of the \(\epsilon\)-SVR with that degree of precision.}. Compared to standard robust linear regression approaches, \autocite{newman_etal15} exhibits the better performance of SVR methods with \enquote{spillover effects} (see \Cref{sec:conclusion-limitations}), enabling them to integrate more closely related cell types in their analysis while providing a more robust and explainable model.

In practice, CIBERSORT implements the $nu-$ SVR approach with the \href{https://cran.r-project.org/web/packages/e1071/e1071.pdf}{svm} function from R package \texttt{e1071} (\autocite{meyer_etal21}).
CIBERSORT additionally provides a standalone web application, and relevant purified signatures. The most popular is the LM22 profile, a meta transcriptomic collection of 6 studies of 22 distinct immune cell types (see \Cref{sec:deconv-pipeline}). 
The ImmuCC algorithm (\autocite{chen_etal17}) harnesses the implementation from CIBERSORT algorithm, with a new reference signature aggregating 25 cell types and tailored for murine deconvolution.

\subsubsection{Correcting the Uncoupling Between RNA and Cytometry Fractions}

It appears that most of the existing deconvolution algorithms estimate the fraction of mRNA coming
attributable to each cell type, rather than the underlying cell proportion itself. In other words, they assume \emph{homogeneous} cell populations, e.g. they consider that each cell subtype exhibits the same RNA library depth (\autocite{sosina_etal21}). However, in real-world settings, this premise usually does not hold, for both technical and biological reasons. For instance, the RNA extraction efficiency may depend on the cell type, and its survival capacity to the lysis and extraction phase.
Once the average production of total transcriptomic expression has been estimated (or phsyically measured), it becomes feasible to subsequently re-normalise the inferred cellular transcriptomic ratios, such that they align with the anticipated, biologically interpretable cellular ratios (see \Cref{eq:normalisation-uncoupling-rna-fraction}):

\begin{equation}
\hat{p^*_j} = K \frac{\hat{p_j}}{r_j}, \quad K= \frac{1}{\sum_{j=1}^J \frac{\hat{p_j}}{r_j}}
\label{eq:normalisation-uncoupling-rna-fraction}
\end{equation}
with \(r_j\) the average number of transcripts extracted per cell type, and \(K\) the normalisation constant.

Post-correction of this uncoupling is accounted in \autocite{racle_etal17} and \autocite{finotello_etal19} studies, with direct measures of the total expression of cell subtypes, as quantified with RNAeasy mini kit (Qiagen)
and he Proteasome Subunit Beta 2, respectively \footnote{In the back-end, they utilise the expression of the \textit{housekeeping genes} as a surrogate variable of the absolute number of transcripts produced by the cell population}.

When direct measures are not available, the MMAD (microarray microdissection with analysis of differences, \autocite{liebner_etal14}) proposes an iterated approach for estimating the coefficient extraction efficiency, \(r_j\). Yet, the regression framework is not anymore linear, and the new cellular estimate is computed using a non-linear conjugate gradient search algorithm.

\subsubsection{Linear Regression Approaches with Explicit Unit-Simplex Constraint}

All the previously described algorithms do not explicitly integrate the unit-simplex constraint \Cref{eq:simplex-constraint} during the estimation process, and re-normalise instead, posterior to the estimation, the inferred ratios.

The NNLS (Non Negative Least Squares) estimate relies on the Lawson Hanson algorithm \autocite{haskell_hanson81}, and its output is often provided as a reference in most review papers benchmarking deconvolution algorithms (\autocite{sturm_etal19}, \autocite{jin_liu21}). The \href{https://rdrr.io/cran/limSolve/man/nnls.html}{nnls} function from the R \texttt{limSolve} package can be used to solve this optimisation problem.

The Least Squares with Equality and Inequality Constraints (LSEI) generalises this approach by enforcing both non-negativity and sum-to-one constraints. The \emph{lsei} function in R, from limSolve package, can be used to
solve the corresponding optimisation problem. The Matlab \emph{lsqlin} function, returning the same output as \emph{lsei}, is used by the Bioconductor package  \texttt{DeconRNASeq} (\autocite{gong_etal11}, \autocite{gong_szustakowski13}) . 

Both algorithms belong to the class of \emph{QP} (quadratic programming), which aims at optimising a system of linear, convex functions, with a guaranteed unique solution.

\subsubsection{Regularised linear regression}
\label{para:regularisation-approach}

When the number of cell types \(J\) exceeds the number of transcripts \(G\), the deconvolution problem stated in \Cref{eq:linear-deconvolution} is \emph{undetermined}, with potential infinite set of solutions verifying the set of $G$ equations. Several regularised linear approaches have been implemented to deal specifically with problems where the number of unknowns exceeds the number of variables (see Definition A.7).

 The DCQ algorithm \autocite{altboum_etal14} uses in particular the \textbf{Elastic Net} regularisation, a compromise between the L1 and L2 penalties proposed by the Lasso and Ridge methods. In R, the \emph{glmnet} \autocite{friedman_etal11} offers a straightforward and versatile implementation of the method. 
 The benchmark study led by \autocite{jin_liu21} exhibits the reduced performance of deconvolution methods applying these regularised approaches. However, a comprehensive analysis of the settings used to conduct the benchmark study show that they somehow miss the point: penalised linear regression approaches are not intended to retrieve the cell ratios of a given biological sample, but rather retrieve the optimal \textit{support} of cell populations that induce transcriptomic variations from a biological state to another. Implicitly, these methods assume that the proportions of most cell populations do not vary over time.

\bigskip
To illustrate the point, DCQ has been used to identify the dynamical evolution of immune cell ratios during influenza infection. Indeed, dozens of immune cell types coordinate their efforts to maintain tissue homeostasis. Precisely, DCQ studied the evolution dynamics of up to to 213 immune cell subpopulations in mice lungs for ten time points and retrieve significant changes in 70 immune cell type ratios. 

Two years after, the ImmQuant package \autocite{frishberg_etal16} offers a user-friendly tool for inferring immune cells in both human and mice organisms. The pipeline includes automatic data import and cleansing, selection of the marker genes, deconvolution of the biological samples provided and visualisation of the output. 

\subsection{Probabilistic-based approaches}

The second family of methods for inferring cellular ratios from purified reference profiles utilises probabilistic models to capture the generative process underlying the bulk expression production. Interestingly, these approaches naturally address the unit-simplex constraint (\Cref{eq:simplex-constraint}), provide a more accurate representation of the discrete nature of transcript counts and can even account for an unknown cell population or individual variations of the gene expression. In particular, these approaches accurately reproduce the commonly observed correlation between the mean and the variance of the gene expression (\autocite{lobenhofer_etal08}).

Since a large number of parameters might be introduced in these models, it is common practice to represent the conditional independence relating them using a directed acyclic graph and the homogenised notation illustrated in \Cref{subfig:deconvolution-dag-legend}.

\subsubsection{Discrete probabilistic approaches}
\label{para:deconv-proba-discrete}

Latent Dirichlet Allocation (LDA) is a straightforward approach to model abundances (see also \autocite{blei_etal03} and Definition A.8). The NNML (Non-negative maximum likelihood model) algorithm, by \autocite{qiao_etal12}, extends the frequentist LDA model adopting a Bayesian approach.
Precisely, the prior distribution of the cell ratios is modelled by a symmetric Dirichlet distribution. This kind of distributions exhibits several advantages: it naturally endorses the unit-simplex constraint \Cref{eq:simplex-constraint} and streamlines the integration of prior knowledge, such as equibalanced hypothesis or inclusion of cytometry measures \footnote{To note, the Beta distribution is a variant of Dirichlet distribution with two-component mixtures, used as prior for binomial distributions.}

Extensions of the NNML algorithm introduce generative models that relax controversial assumption, such as the completeness (no unknown cell population) or the validity (no sample-specific variations of the purified signatures) of the reference profile. However, these probabilistic frameworks often require \textbf{regularisation} strategies, classified as \enquote{hard} and \enquote{soft} constraints, to ensure problem \textit{identifiability}. Practical regularisation strategies often rely on strong constraints and assumptions about the distribution of purified expression profiles. They must balance the trade-off between introducing too much bias and risk overfitting, or insufficiently define the problem and suffer from \textit{ill-conditioned} modelling.

\bigskip
To that end, the \texttt{ISOLATE} algorithm (\autocite{quon_morris09}) assumes that the expression profile of any gene of the unknown cell type can be rewritten as the expression of one of the cell types already described, up to an additional multiplicative perturbation described by an uninformative Gamma prior. In a tumoral context, this constraint can be interpreted as a change of gene expression induced by heterotypic tumoral conditions, on an unique cell population subset, termed CSO in the paper (cancer site of origin). 
The basic framework described above has been extended in the \texttt{ISOpure} algorithm (\autocite{quon_etal13}). Unlike the naive approach, \texttt{ISOpure} not only computes a shared cancer profile common across all samples but also refines it to incorporate sample-specific variations in tumoral expression.
However, the CSO assumption only holds if the mutations concern only one cell line, an assumption that usually does not hold in intricate TMEs, wehre both tumoral and normal cell lines expression are impacted by the clonal growth.

Accordingly, the \(\text{NNML}_\text{np}\) algorithm (\autocite{qiao_etal12} and \Cref{subfig:deconvolution-discrete-models}) assumes instead that the transcriptomic profile of the unknown cell type can be rewritten
as a potential convex combination of all (possibly a subset) the included cell populations. Biologically, this approach hypothesises that the tumoral part of the sample is not a new cell line, but rather a mixture itself of the original cell populations, whose expression has been altered upon tumoral mutations, or changes induced by the new conditions of the medium.  Their approach is nonetheless hindered by the stringent regularisation assumption that the perturbation factor for a given gene is the same across cell populations.

The PERT algorithm (\autocite{qiao_etal12} and \Cref{subfig:deconvolution-discrete-models}) relaxes the strong assumption that the purified cell expression profiles are representative of the expression profiles of the mixture.  Specifically, the vector representing the expression profile of a cell population is altered through a multiplicative perturbation factor \(\boldsymbol{\rho}_G\), which is gene-specific and sampled from a non-informative Gamma distribution with an average value of 1.
\bigskip

TEMT (Transcript Estimation from Mixed Tissue samples, \Cref{subfig:deconvolution-discrete-models} ), by \autocite{li_xie13}, harnesses directly the reads (sequence of nucleotides) themselves, instead of raw RNA-Seq counts. This approach enables to account for multiple transcripts resulting from \emph{alternative splicing} (\autocite[Chap 14]{campbell_etal20}) and technical biases issued from read sequencing itself \footnote{Technical artefacts in RNA-Seq encompass length, positional and amino bias. For instance, longer transcripts may yield more counts (\enquote{effective length}), while sequence-related biases include over-transcription around transcript ends.}. The methodology is thus particularly relevant for decomposing, and correcting technical artefacts from relevant biological signal, and can be used as an alternative normalisation method for making samples comparable, regardless of the sequencing platform used.

This approach uniquely incorporates technical artefacts into the deconvolution process, addressing the assumption made by other methods that input data has been corrected for such noise. Additionally, it estimates an unknown cell profile, in a process similar to the \(\text{NNML}_\text{np}\) approach.

\bigskip

The complexity of the likelihood or the posterior function requires specific optimisation methods to retrieve the relevant parameters: PERT and NNML uses a conjugate gradient descent algorithm, while TEMT and the ISOLATE algorithm utilise a variational online EM \autocite{dempster_etal77}.
Since diverse regularisation strategies do not address the same biological constraints, and often require different optimisation strategies, \autocite{quon_morris09} suggests to systematically benchmark the method against manually annotated tumours, as evaluated by pathologists. 

\subsubsection{Continuous probabilistic approaches}
\label{subsubsec:continous-probalistic-approaches}

\begin{figure}
\centering
     \begin{subfigure}[p]{0.45\textwidth}
     \centering
         \includegraphics[width=0.6\textwidth]{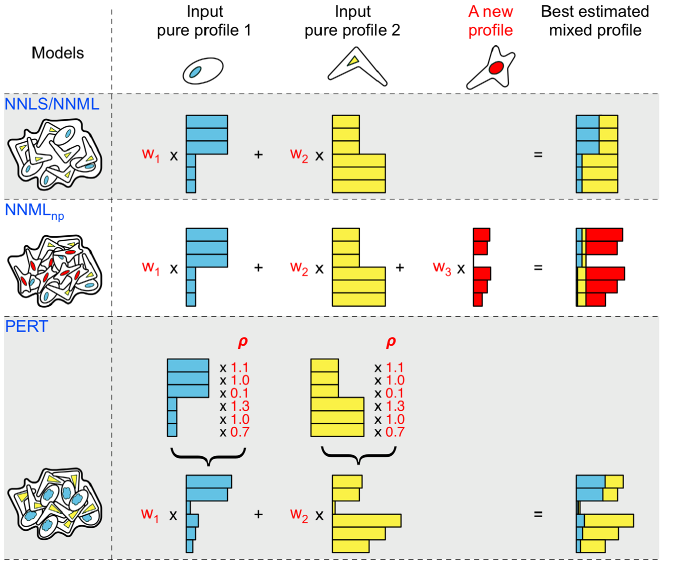}
         \label{subfig:pert-deconvolution}
         \caption{\textbf{To each biological requirement, its suited probabilistic model.} The non-negative least squares model (NNLS) and the non-negative maximum likelihood model (NNML) can only predict proportions of pre-specified reference populations. In scenario ii), the non-negative maximum likelihood new population model (NNMLnp) can additionally account for an unknown cell population, while in scenario iii) the perturbation model (PERT) can integrate sample-specific variations. Reproduced from \autocite[Fig. 1]{qiao_etal12}.} 
     \end{subfigure}
     \hfill
     \begin{subfigure}[p]{0.45\textwidth}
     \centering
         \includegraphics[width=0.8\textwidth]{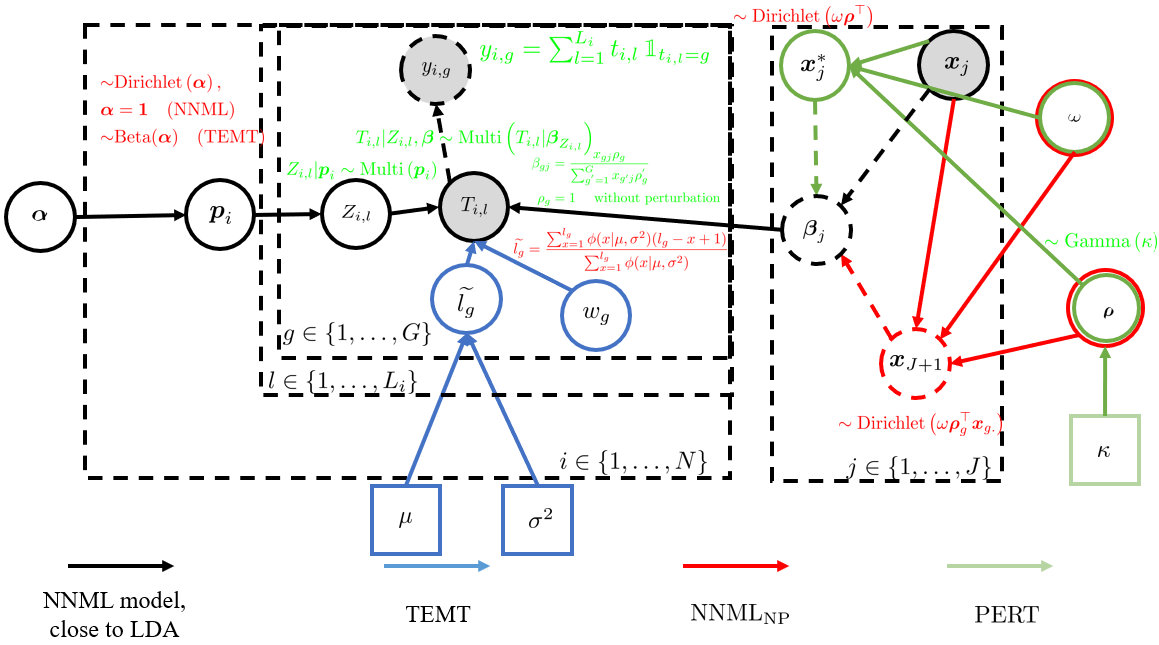}
         \label{subfig:deconvolution-discrete-models}
         \caption{\textbf{DAGs of the generative model described in \cref{para:deconv-proba-discrete}.} All the discrete probabilistic models derived from the LDA generative framework. This DAG notably encompasses, using different colour notations, the NNML, $\text{NNML}_\text{NP}$ and PERT algorithms (\autocite{qiao_etal12}), along wihth the TEMT model \autocite{li_xie13}.} 
     \end{subfigure}     
     \vfill
     \begin{subfigure}[p]{0.45\textwidth}
     \centering
         \includegraphics[width=\textwidth]{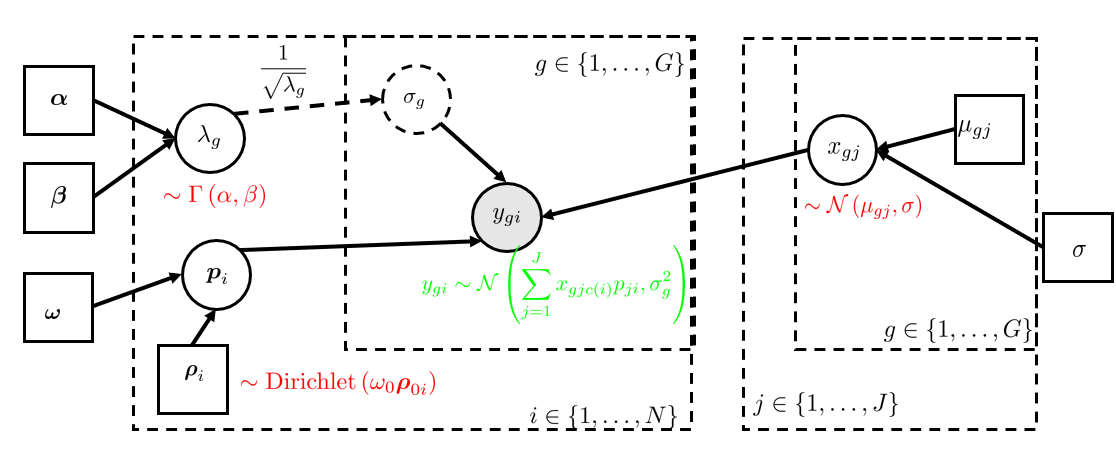}
         \label{subfig:deconvolution-demixt}
         \caption{\textbf{Graphical representation of the Demix(T) (\autocite{ahn_etal13} and \autocite{wang_etal18}) probabilistic model.} }
     \end{subfigure}   
     \hfill
     \begin{subfigure}[p]{0.45\textwidth}
     \centering
         \includegraphics[width=\textwidth]{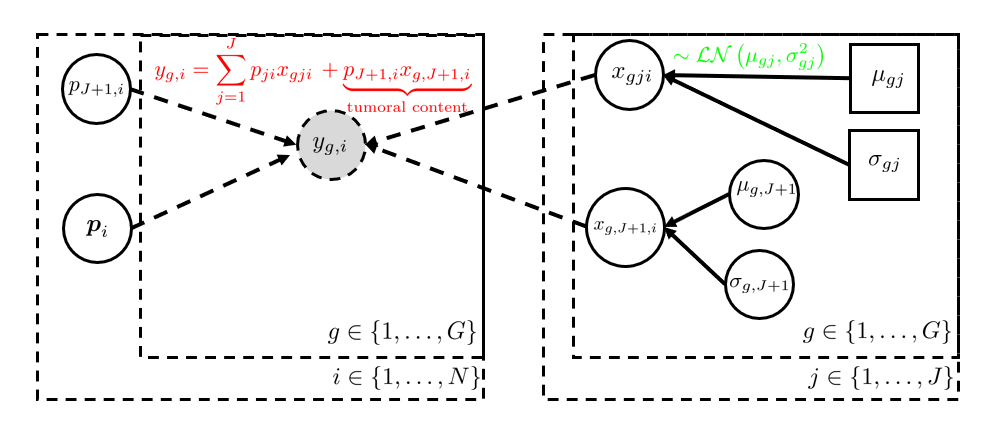}
         \label{subfig:deconvolution-Dsection}
         \caption{\textbf{Graphical representation of the Dsection \autocite{erkkila_etal10} probabilistic model.} }
     \end{subfigure}   
     \vfill
     \begin{subfigure}[p]{0.6\textwidth}
     \centering
         \includegraphics[width=\textwidth]{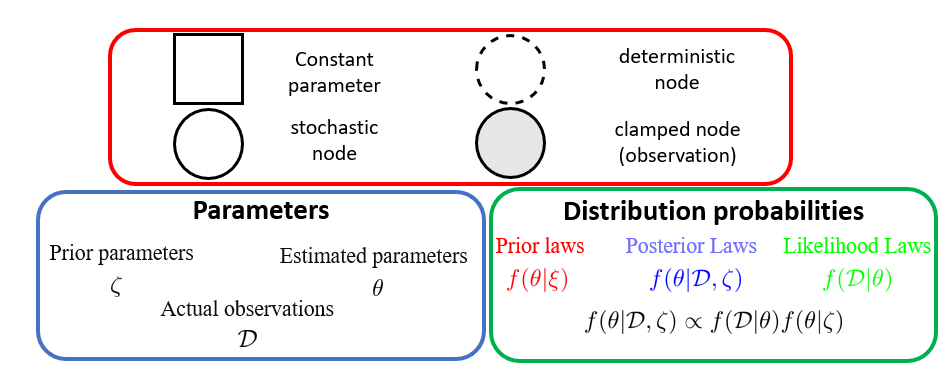}
         \label{subfig:deconvolution-dag-legend}
     \end{subfigure}   
     \caption{\textbf{Partial probabilistic models to infer cellular ratios}.We follow the \href{https://revbayes.github.io/tutorials/intro/graph_models.html}{RevBayes} convention to homogenise indexes and parameters across a set of generative models. Notably, the \textit{likelihood} density functions describing the distribution of the observations, are in green colour while the prior distributions of the parameters to estimate are in red colour.} 
\end{figure}

      The \texttt{Demix} generative model, by \autocite{ahn_etal13}, and its direct \texttt{DemixT} extension,by \autocite{wang_etal18}, infer the proportion and expression profile of the tumoral content, in a two and three-component mixture, respectively.    
     Briefly, Demix(T) models the distribution of the bulk expression for each gene as a convolution (sum of independent variables) univariate log Normal distributions (see \Cref{subfig:deconvolution-demixt}), each purified profile parametrised by its own parameters, inferred prior to the study. 
      For the sake of comparison, a generative model based on a convolution of Normal distributions is also compared to the log Normal approach. This model streamlines the estimation process as a closed-form can be derived for the log-likelihood. However, the \(\log_2\)-transformation required to endorse the assumptions of the model is likely to disrupt the fundamental linearity deconvolution assumption (\Cref{eq:linear-deconvolution}).
     
     Modelling the mixture problem as a convolution offers several advantages, including the elimination of a residual error term to account for the stochasticity of the resulting bulk profile, and the utilisation of distributions that accurately depict  the inherent compositional characteristics of RNA-Seq datasets.
     
     However, no explicit form for the convolution of \(\log_2\)-normalised variables is known, and an iterated conditional modes-like (\autocite{besag86}) approach \footnote{The parameters are iteratively maximised, conditioned on the current updated value of the remaining subset of parameters, rather than simultaneously} is used to maximise the log-likelihood of the resulting generative model:
     \begin{itemize}
         \item The unknown general parameters of interest (cellular proportions and mean and variance of the tumoral profile), are determined by maximising the log-likelihood of the generative model depicting the convolution, conditioned on the previously known mean and variance for healthy cell populations. Since the closed form of the log-likelihood is not known for a convolution of log-Normal, it is approximated through numerical integration (not needed with a convolution of Normal distributions), and the MLE is obtained using a \textit{Nelder-Mead} procedure.         
         \item In a second time, tumoral profiles are estimated by plugging-in the parameters estimated in the previous step. With a two-component model, the unit-simplex constraint (\Cref{eq:simplex-constraint}) and the fundamental linear deconvolution assumption (\Cref{eq:linear-deconvolution}), only one degree of freedom, or unknown, namely the tumoral content, must be inferred (see \autocite[Eq.1]{ahn_etal13}).
     \end{itemize}

     \bigskip
     \bigskip
     
  \autocite{erkkila_etal10} implements instead a Bayesian framework, \texttt{Dsection} (see \Cref{subfig:deconvolution-Dsection}, in which the bulk expression of each gene in each sample, \(y_{gi}\), follows a Normal distribution whose parameters are stochastic variables rather than point values. For instance, the distribution of the inverse of the variance,  referred to \emph{precision} in the paper, is modelled by a Gamma distribution.

  The posterior distribution of individual cell-specific expressions and bulk gene variances is identifiable to known density distributions (\textit{conjugate} priors). However, the posterior distribution of cellular ratios lacks a known density distribution due to the intractable integration of the normalising constant. The Metropolis-Hasting algorithm is employed to sample this posterior distribution, which is only known up to a normalising constant, while Gibbs sampling is used to retrieve simultaneously the joined posterior distributions of the whole set of parameters composing the generative model.
  Note that in opposition to the Demix(T) approach (\autocite{ahn_etal13}), the variance of the bulk expression is uncoupled to the individual variance of the purified cellular profiles.

\section{Marker-Based Approaches: Pathway Enrichment Analysis and Hyper-Geometric Scores}
\label{subsec:marker-based-approaches}

Some deconvolution algorithms simplify the estimation process by adopting a marker-based paradigm. The definition of \enquote{markers} genes has gradually broadened, from designing genes uniquely expressed in a cell a population to include genes comprehensively expressed in one cell type relatively to other cell groups. 
Marker-based relied historically on strong definitions of \textit{marker} genes   (\autocite{gosink_etal07}, \autocite{clarke_etal10}), however, nowadays, \emph{weak} markers approaches are favoured (markers are only required to be consistently over-expressed in a given cell population), since they also enable to delineate closely related cell types. 

These markers can be derived through either knowledge-driven approaches (\autocite{angelova_etal15}, \autocite{rooney_etal15}) or data-driven methods \autocite{chikina_etal15}, \autocite{becht_etal16}, \autocite{zhang_etal17}. The initial data-driven strategy for identifying marker genes involved identifying genes whose mean expression value in a give cell population consistently exceeded the expression value measured across other cell types (\autocite{shoemaker_etal12}, \autocite{chikina_etal15}).
More robust statistical approaches, evaluating the relevance of selected markers through the computing of empirically estimated\emph{p}-values, have been developed since then, ranging from SNR (signal-to-noise) ratios \autocite{becht_etal16}, to the F-statistic (\autocite{wang_etal10}) through the Gini index (\autocite{zhang_etal17}). 

\bigskip
Integrating the definition of a gene marker into the fundamental presumption of linear deconvolution simplifies framework \Cref{eq:linear-deconvolution}) into \Cref{eq:marker-relation}: 

\begin{equation}
    \begin{split}
y_{\forall g \in \widetilde{G_j}}&= \sum_{j^{'}=1}^{J} x_{g j^{'}} \times p_{j^{'}} = x_{gj} p_{j}, \\ &\text{ since by definition } x_{gj^{'}} =0, \forall j^{'} \neq j \\
\begin{pmatrix}
\boldsymbol{y}_{\widetilde{G_1}} \\
\boldsymbol{y}_{\widetilde{G_2}} \\
\vdots  \\
\boldsymbol{y}_{\widetilde{G_J}} 
\end{pmatrix}
&=
\begin{pmatrix}
\boldsymbol{x}_{\widetilde{G_1},1}  &\ldots & \boldsymbol{0}\\
\boldsymbol{0} & \boldsymbol{x}_{\widetilde{G_2}, 2}   & \boldsymbol{0}\\
\vdots & \ddots & \vdots\\
\boldsymbol{0} & \ldots & \boldsymbol{x}_{\widetilde{G_J}, J}
\end{pmatrix}
\times
\begin{pmatrix}
p_{1} \\
p_{2} \\
\vdots  \\
p_{J} 
\end{pmatrix}
\end{split}
 \label{eq:marker-relation}
\end{equation}

with the following notations:
\begin{itemize}
    \item $\widetilde{G}=\{ 1, \ldots, G\}$ is the set indexing the total number of genes selected in the signature matrix (we introduce the tilde as a shorthand indicator for a set).
    \item $\widetilde{G_j} \subset \widetilde{G}$ is the subset of genes expressed uniquely in cell population $j \in \widetilde{J}$
    \item We additionally assume the unique existence of a \textit{partition} \(\widetilde{G}\), shared across samples, such that $\widetilde{G_j} \cap \widetilde{G_l} = \varnothing, \, \forall (l, j) \in \widetilde{J}, \, l\neq j$ and $\bigcup_{j=1}^J \widetilde{G_j} = \widetilde{G}$.
    \item We introduce the shorthand \(\boldsymbol{y}_{\widetilde{G_j}}\) and $\boldsymbol{X}_{\widetilde{G_j}, j}$ to respectively denote the measured expression of the market set $\widetilde{G_j}$ in the bulk mixture, and its respective expression in the purified cell population $j$.
\end{itemize}

If \cref{eq:marker-relation} holds, the bulk expression associated to a gene marker set is proportional to the expression of the cell population associated to this marker, the multiplicative constant being the ratio associated to this cell type, $p_j$. 

However, as already specified in \Cref{sec:reference-based}, the presence of technical noise or intrinsic biological stochasticiy usually renders the system of equations inconsistent. Assuming the same framework detailed in \Cref{par:linear-modelling-cellular-ratios}, the Normal Equations give the following OLS solution (\Cref{eq:marker-solution}):
\begin{equation}
    \hat{p_{j}} = \frac{1}{|G_j|}\sum_{g \in \widetilde{G_j}} \frac{y_{g}}{x_{gj}}    
 \label{eq:marker-solution}
\end{equation}
with \(|G_j|\) the module, namely the number of genes composing the marker set of a cell population. 

\bigskip
Once specific markers for each population have been identified, the estimation of cellular ratios relies either on \textit{abundance score} (see \Cref{para:abundance-scores}) or \textit{enrichment score} (see \Cref{para:enrichment-scores-KS} and \Cref{para:enrichment-scores-chi2}). 

\subsection{Abundance scores}
\label{para:abundance-scores}

Historical endeavours, by \autocite{gosink_etal07} and \autocite{clarke_etal10}, assume the strong definition of a marker (\cref{subsec:marker-based-approaches}) holds, and the cellular ratios that were returned correspond to the estimates given in \cref{eq:marker-solution}. \autocite{clarke_etal10} only differed by the addition of a \emph{link function}, precisely a \emph{log2} transformation to reduce the noise bias associated to small ratio values, applied to the bulk and purified profiles. 

Later, the MCP (Micro-environment Cell Populations)-counter, by \autocite{becht_etal16}, adopts a weak marker paradigm, and replaces the abundance score given in \Cref{eq:marker-solution}, by the geometric mean of the genes characterising a given cell population (\cref{eq:MCP-score}):

\begin{equation}
    \ES(\widetilde{G_j} \in \widetilde{G}) = \left(\prod_{g \in \widetilde{G_j}} y_j\right)^{1/\left|\widetilde{G_j}\right|} \propto p_j
 \label{eq:MCP-score}
\end{equation}

\subsection{Enrichment scores, based on KS metric}
\label{para:enrichment-scores-KS}

Most of the methods computing an enrichment score rely on a variant of the weighted enrichment-based method named ssGSEA, for single-sample gene set enrichment analysis (\autocite{subramanian_etal05} and \autocite{barbie_etal09}). The computation of enrichment scores, based on the Kolmogorov–Smirnov metric, is reported in Definition B.1, while its main limitations.

\autocite{yoshihara_etal13} implements the ESTIMATE metric to compute immune and stromal enrichment scores in tumoral samples. Te best link function coupling the purity score (proportion of tumoral cells) with the ESTIMATE measure was computed with the \href{Eureqa}{https://en.wikipedia.org/wiki/Eureqa} software.
\autocite{aran_etal15} implements an extension of this method integrating orthogonal modalities. Precisely, the tumour purity score is computed from four distinct sources: the ESTIMATE score itself , ABSOLUTE (quantify the proportion of cancer cells based on the number and location of somatic copy-number mutations), LUMP (correlation between the degree of methylation and the tumour proportion) and immunehistochemistry image analysis. 

\autocite{rooney_etal15} and \autocite{angelova_etal15} uses GSEA-based metrics to compute the tumoral activity and relate it to mechanisms involved in immune tumour resistance. \autocite{angelova_etal15} notably demonstrates the co-existence of two kinds of tumoural environments, distinguishing hypermutated tumours showing upregulation of immunoinhibitory molecules from non-hypermutated and stagnant tumours, enriched with immunosuppressive cells. 

\autocite{senbabaoglu_etal16} infers gene markers for 24 distinct cell populations in 19 cancer types. With these enrichment scores, they demonstrate that the over-expression of Th17, CD8+ and Tregs increases chances of survival, while strong activity of Th2 cells is correlated with a negative prognostic.

Ultimately, the \texttt{xCell} algorithm, by \autocite{aran_etal17}, claims to identify up to 64 distinct cell types, including immune and stromal ones, derived from a compendium of 1822 purified transcriptomic cell lines. \textit{Calibration}, using a power link function to couple abundance scores with true cell ratios, and reduction of the multi-collinearity of the signature matrix to avoid \enquote{spillover} effects, underlie the originality, and robustness of the method.

Finally, \href{https://icbi.i-med.ac.at/software/timiner/doc/}{TIminer}, by \autocite{tappeiner_etal17}, is a free Docker pipeline, aggregating the marker sets of \autocite{aran_etal17}, \autocite{angelova_etal15} and \autocite{charoentong_etal17}. It was initially designed for estimating the proportion of infiltrated immune cell types, along with neoantigen prediction and tumour immunogenicity.

\subsection{Enrichment scores, based on alternative metrics}
\label{para:enrichment-scores-chi2}
Alternative strategies can be employed to compute enrichment scores, such the hypergeometric test (see Definition B.2).

\autocite{bolen_etal11} implements SPEC (Subset Prediction from Enrichment Correlation) to predict which cell population is more likely to contribute to an observed change in the gene expression, based on Pearson correlation. SPEC notably demonstrates that the main resistance mechanism of the gold-standard treatment against Hepatis C was the cross-interaction between the myeloid cells and the anti-interferon therapy.

\autocite{shoemaker_etal12} uses the $z$-score (negative log10 of \emph{p}-value), resulting from a Fisher's exact test. 

The Bioconductor package \href{https://www.Bioconductor.org/packages/release/bioc/html/BioQC.html}{BioQC}, by \autocite{zhang_etal17}, computes abundance scores by evaluating the relevance of median differential expressions with a non-parametric \emph{Wilcoxon-Mann-Whitney} test.

\bigskip

In conclusion, marker-based methodologies provide abundance scores that are only proxy of relative cellular ratios. \autocite{aran_etal17} and \autocite{yoshihara_etal13} attempt to mitigate this issue, by learning a link function coupling these two features. Overall, these restrictions render marker-based methods impractical for intra-sample comparisons, in contrast to the signature-based methods, discussed in previous \Cref{sec:reference-based} (see also Appendix B.3).

We outline the major categories of deconvolution algorithms used to estimate cell ratios in a heterogeneous biological sample in \Cref{fig:classification-deconvolution}:

 \begin{figure}[H]
          \centering
          \includegraphics[width=0.8\textwidth]{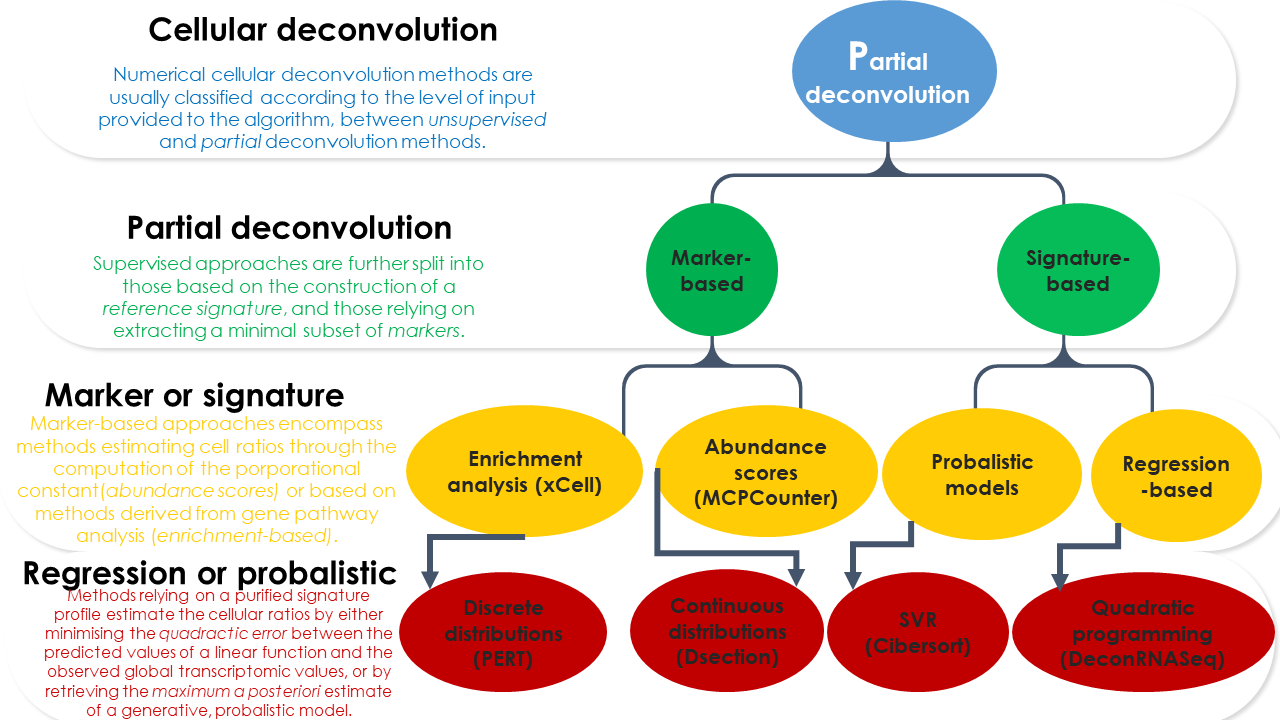}
          \caption{General classification of partial-based deconvolution algorithms.}
          \label{fig:classification-deconvolution}
\end{figure} 

\section{Reference-Free Approaches: Simultaneous Deconvolution of Cell Fractions and Purified Expression Profiles}
\label{subsec:unsupervised-and-reference-free-deconvolution-methods}

Complete deconvolution algorithms attempt to simultaneously estimate both the proportions and the pure expression profile of cell types \autocite{shen-orr_gaujoux13} from the bulk profile alone, namely minimising the following quantity (\Cref{eq:unsupervised-deconvolution}):

\begin{equation}
\left(\hat{\boldsymbol{P}}, \hat{\boldsymbol{X}}\right) = \arg \min_{ \boldsymbol{P}, \boldsymbol{X}} \, \{ \left|\boldsymbol{Y} - \boldsymbol{X} \times \boldsymbol{P}\right| \} \quad
\boldsymbol{Y} \in \mathbb{R}^{G \times N}_+ , \, \boldsymbol{X} \in \mathbb{R}^{G \times J}_+, \,\boldsymbol{P} \in \mathbb{R}^{J \times N}_+
 \label{eq:unsupervised-deconvolution}
\end{equation}

Without further information, the system of equations described in \Cref{eq:unsupervised-deconvolution} is \textit{undetermined}, having either an infinite set of solutions or no one at all. Hence, the identifiability of the unsupervised deconvolution problem require strong assumptions on the distribution.

\subsection{Unsupervised approaches}

\autocite{venet_etal01} proposes the first version of a reference-free approach, inspired from Gaussian mixtures, to deconvolve colon cancer samples, from which two clusters, on a total of four identified, could be labelled with strong evidence as hematopoietic and fibroblast cells. 
\autocite{venet_etal01} also demonstrates that the marker-based assumption (see \Cref{subsec:marker-based-approaches}) is a necessary condition for the existence and uniqueness of the system of equations  (\Cref{eq:unsupervised-deconvolution}).

Repsilber and colleagues then extended the method proposed by \autocite{venet_etal01}, by solving \Cref{eq:unsupervised-deconvolution} using a Non-Negative Matrix Factorisation algorithm. NMF notably guarantees that both $\boldsymbol{X}$ and $\boldsymbol{P}$ are strictly non-negative  (see Definition C.1 and \autocite{repsilber_etal10}), as reported in \Cref{eq:unsupervised-deconvolution-nmf1}:

\begin{equation}
    \begin{split}
        &\min_{\boldsymbol{P}, \boldsymbol{X}} \lVert \boldsymbol{Y} - \boldsymbol{P}\boldsymbol{X} \rVert_F^2 \\
        &\text{ subject to the non-negativity constraints: } \\
        &\boldsymbol{P} \geq 0, \, \boldsymbol{X} \geq 0
    \end{split}
 \label{eq:unsupervised-deconvolution-nmf1}
\end{equation}

Variants of the NMF approach were used in UNDO, by \autocite{wang_etal15} and CAM, by \autocite{wang_etal16}, methodologies. The Convex Analysis of Mixtures (CAM) enforces both the non-negativity of the outputs returned, and the unit-simplex constraint \Cref{eq:simplex-constraint} for the ratios. Precisely, these convex geometry-based methods project the resulting bulk expression matrix $\boldsymbol{Y}$ into a $J$-dimensional \emph{polytope}, whereby each cell population profile forms a convex hull whose vertices are the marker genes of the so-called cell population. The final set of convex solutions are the ones covering the most precisely the facets of the convex hulls derived from the bulk profile.
\href{https://rdrr.io/bioc/CAMTHC/}{CAMTHC}, by \autocite{chen19}, for Convex Analysis of Mixtures for Tissue Heterogeneity Characterisation, and \textit{CAMfree}, by \autocite{jin_liu21}, are both R package implementing the CAM methodology.

\subsection{Semi-supervised approaches integrating prior information}

Since then, semi-supervised approaches, coupling partial prior knowledge of markers associated with a cell type with numerically inferred \texttt{de-novo} molecular markers, enable to increase the identifiability of the problem by reducing the set of possible solutions. Semi-approaches directly extending \autocite{wang_etal16} have been implemented in R, as packages \textit{CAMmarker} and \textit{CellMix} \footnote{\textit{CellMix}, by \autocite{gaujoux13}, benchmarks a whole set of deconvolution methods, in particular, \textit{ssKL} and \textit{ssFrobenius} that solve optimisation problem \Cref{eq:unsupervised-deconvolution-nmf1} by minimising the KullBack Leibler divergence and the Frobenius norm, respectively.} The usual approach to integrate prior information is to constrain all input values of the purified expression profile to zero, except whether the gene has formally been associated with a cell population.

Closely related is the semi-CAM approach, by \autocite{dong_etal20}. In details, the semi-CAM approach is a two-step estimation procedure; first, it identifies the final gene partition for the deconvolution process, assigning each unlabelled gene to its most probable cell type, given the already identified marker genes. To achieve this, it enhances the $k$-means clustering employed by the CAMfree approach, whereby the initial centroids are the vertices covering the most the convex hulls, by incorporating known marker information into the cluster centre construction.
Whenever known marker genes for partially described cell types are available, \autocite{dong_etal20} demonstrates that the semi-CAM method outperforms the unsupervised historical CAMfree method. 

\bigskip

The Digital Sorting Algorithm (DSA, \autocite{zhong_etal13}), is another semi-supervised approach, adopting a EM-like approach. Precisely, the cellular ratios and the purified expression profiles are iteratively estimated, conditioned on the current update of the remaining parameters, until convergence. Prior information can easily be integrated as initial values for either cellular ratios or purified expression profiles. However, the identifiability of the problem still requires the marker assumption. 

Overall, all the methods described in this section are much more sensitive to the quality of data provided, especially when no prior information is provided. 

\section*{Outline of the Cellular Deconvolution Procedure}
\label{sec:deconv-pipeline}

The estimation of the composition of a biological sample is only one of the steps composing the deconvolution framework.  In the remainder of the text, we define as \textit{pipeline} this whole process, ranging from the pre-processing and collection of purified profiles to the downstream analyses, while the term \enquote{algorithm} only refers to the estimation stage itself.

A standard cellular deconvolution pipeline typically involves the following main steps:

\begin{enumerate}
\item 
\textbf{Data Preprocessing and Marker Gene Selection:} This step (see stage 1, in \Cref{subfig:deconvolution-pipeline}) involves the formatting of gene expression profiles obtained through RNA-seq or microarray, ranging from quality control to data transformation transformation and normalisation, and the removal of unwanted batch effects induced by technical artefacts. 

\item \textbf{Construction of purified signature matrices} Partial methods inferring cell ratios requires an additional step consisting of identifying and characterising a subset of genes, able to delineate all the cell populations ought to compose the mixture. This step is illustrated in \Cref{subfig:deconvolution-pipeline}, part 2.

\item \textbf{Parameter Estimation:} This step refers to the deconvolution algorithm itself (stage 3, \Cref{subfig:deconvolution-pipeline}). The type of tissue or/and organism to deconvolve along with the objective biological goal guide the final choice of the algorithm used. 

\item  \textbf{Evaluate the output:} This step involves the formulation of statistical tests to assess the presence of a cell population within the sample (intra-sample comparison) or to compare two cell fractions across different biological conditions (inter-sample comparison). Surprisingly, there is a notable absence of robust and widely accepted methods proving theoretically the consistency and precision of the outputs returned by most deconvolution methods. Alternatively, it is possible to benchmark the performance of a new deconvolution algorithm against gold-standard deconvolution methods and against cytometry data. 

\item \textbf{Visualisation and biological interpretation: }
Ultimately, various visualisations and expert validations play a pivotal role in verifying the precision and biological relevance of the algorithm in deciphering disease mechanisms, or providing new biomarkers (see stage 4, in \Cref{subfig:deconvolution-pipeline}). 
\end{enumerate}

A practical use case, with the construction and the application of the LM22 signature in conjunction with the CIBERSORT algorithm is reported in \autocite{chen_etal18}.

\bigskip
\bigskip
We now delve into the methods used for selecting the minimal subset of genes, that best discriminate the cell populations included in the deconvolution study. Overall, they fall under the general \emph{feature-engineering} machine-learning concept, which refers to the preprocessing stage that filters irrelevant variables before applying the model \autocite{guyon_elisseeff03}.

Precisely, partial deconvolution methods based on signature profiles (\Cref{sec:reference-based}) typically employ the \enquote{one-vs-all strategy} to identify the minimal set of transcripts consistently expressed in a given cell population, compared to all others. This strategy notably aims to reduce gene expression variance within a given cell type while simultaneously maximising the variance between different cell populations. However, once concatenated, the number of identified markers is still usually intractable to perform deconvolution tasks, and the resulting signature matrix often exhibits strong multicollinearity. Thus, most partial deconvolution approaches integrate an additional step to refine the purified references, which usually enables faster computation, increases the Signal-to-Noise Ratio (SNR) and increases the robustness and reproducibility of the model. 

To select the genes in a global approach, the most common approach, for models based on regression optimisation, relies on optimising the \emph{condition number} of the final reference matrix. In short, the idea is to identify the subset of quantified genes whose combined expression in the transcriptomic expression profile has the smallest condition number.

\section{Main Challenges in the automated quantification of cell populations from RNA sequencing data}
\label{sec:conclusion-limitations}

Several benchmarks have recently been developed to compare the performances of numerical deconvolution methods in relation with the biological objective ( \autocite{sturm_etal19}), the preprocessing protocol chosen to normalise datasets (\autocite{fa_etal20}) or the noise structure and magnitude (\autocite{jin_liu21}).

\subsection{Impact of normalisation techniques}
\label{para:impact-normalisation-deconvolution}
 \autocite{fa_etal20} defines \emph{data normalisation} as the set of techniques to make samples' distribution comparable, including universal scaling methods (min-max, \emph{z}-score, row or column-wise). It also encompasses more specific methods, such as \emph{TPM} or \emph{FPKM}, to account for variations of the library size and depth. On the other hand, \emph{data transformation} refers to the \emph{link function} applied on raw datasets, such that the assumptions underlying the generative model hold. 
 
\autocite{fa_etal20} exhibits that \emph{scaling} methods, such as \textit{row scaling}, or \textit{z-score}, which are used to smooth extreme values, decrease overall the performance of the deconvolution algorithms. In addition, \autocite{fa_etal20} demonstrates that applying log-normalisation leads to suboptimal performances while the best results are reached without transforming the data, conclusions consistent to the findings from \autocite{zhong_etal13}. Indeed, \autocite{hoffmann_etal06} shows that the \emph{log2} transformation, while better guaranteeing the normality requirements on the distribution of the residuals, breaks the fundamental linear assumption (\Cref{eq:linear-deconvolution}).

\autocite{jin_liu21} suggests to apply the same transformations on both the purified signature matrix and the bulk matrix expression, with the best performances obtained with TPM (Transcripts Per Million) normalisation.  \autocite{racle_etal17} indeed suggests that the TPM normalisation, as a \textit{linear mapping}, naturally enforces the unit-simplex constraint \Cref{eq:simplex-constraint}.

\bigskip
Regarding the construction of a signature matrix, \autocite{avilacobos_etal18} emphasises that pre-filtering genes exhibiting the strongest differences between cell types improves the robustness and reproducibility of the algorithm. With LLS-based methods (see \Cref{par:linear-modelling-cellular-ratios}), \autocite{newman_etal15} notably demonstrates the relevance of minimising the \emph{condition number} of the signature matrix, by reducing its multicollinearity (see Appendix D).

To counterbalance technical biases induced by the transcriptomic quantification technology, either RNA-Seq or microarray, some deconvolution methodologies, such as \texttt{CibersortX} (\autocite{newman_etal19}) propose automated batch correction effect with the \href{https://www.rdocumentation.org/packages/sva/versions/3.20.0/topics/ComBat}{ComBat} function, prior to the deconvolution process. Interestingly, \autocite{jin_liu21} demonstrates that Cibersort \autocite{newman_etal15}, CibersortX \autocite{newman_etal19} and MuSiC \autocite{wang_etal19} were less sensitive to the choice of normalisation and sequencing platform, compared to other methods benchmarked.

\subsection{General guidelines for constructing the reference matrix}
\label{para:reference-profile}

\subsubsection{Guidelines for the Selection of Cell Populations for Profiling}
Many deconvolution methods are highly sensitive to the absence of cell subtypes in the reference signature, yielding the best estimates when the reference profile faithfully represents the actual composition of the biological sample \autocite{sturm_etal19}. 

These discrepancies, most pronounced in the absence of closely correlated or orthogonal cellular profiles, lead to the \enquote{spillover} phenomena (\autocite{shen-orr_gaujoux13}, \autocite{fa_etal20}). For instance, \autocite{hao_etal19} demonstrates substantial reduction in estimating the cellular ratios of moncotyes, when myeloid dendritic cells are not included in the reference profile, despite being truly present in the mixture.

On the other hand, \textit{background prediction} refers to erroneous identification of a cell population as being present in a mixture. This issue is even more pronounced with marker-based methods (\cref{subsec:marker-based-approaches}), assuming transcriptomic markers are associated with an unique cell population.
 
Overall, Cibersort \autocite{newman_etal15}, CibersortX \autocite{newman_etal19} and MuSiC \autocite{wang_etal19} are the least sensitive to the presence of undescribed highly-correlated or rare cell types in the mixture (\autocite{jin_liu21}).

 \subsubsection{Guidelines for Phenotype and Tissue Selection in Data Collection}
 To mitigate the recommendations of constructing the most representative cell signature, we should highlight that comprehensive and simultaneous estimation of the whole array of cell populations composing the mixture is usually infeasible. 

 \bigskip
 Firstly, some rare cell types may remain unprofiled, in particular, tumoral profiles are complex to dissect. Tumoral microenvironments display significant variability and plasticity, characterised by distinct mutation patterns, and intra-tumour heterogeneity resulting from the joint presence of diverse tumoral subclones (\autocite{bokil_etal22}). In addition, somatic mutations in native cell lines may lead to the loss of certain markers, posing challenges in defining pro-metastatic immune cell subsets (\autocite{boesch_etal22}), especially for marker-based approaches.  
 
 \textit{TIMERtumour}, by \autocite{li_etal16a} and \textit{EPICabsolute}, by \autocite{racle_etal17}, are computational methodologies specifically tailored to quantify the level of infiltration and contamination of tumoral tissues by immune cells.  
Yet, none of the existing deconvolution methodologies address the intra tumoral heterogeneity, stemming from the potential presence of distinct tumoral subclones (\autocite{yu22}). 

\autocite{racle_etal17} additionally pinpoints that the actual deconvolution solutions for unravelling tumoral heterogeneity are targeted towards decomposition of 
\emph{solid tumours}, rather than "liquid" tumours, such as haematological malignancies (leukaemia).

 \bigskip
 
 Secondly, there is no unique and consistent nomenclature for identifying immune cell subsets, as translating functional insights into reliable phenotypic definitions based on protein markers is challenging (\autocite{aalderen_etal21}). 
It is noteworthy to mention that a suite of R packages, \texttt{ontologyX} \autocite{greene_etal17}, specially tailored to store biological annotations in a structured and tree-like format, have been developed in order to homogenise cell nomenclature with updated ontologies, integrating updated cell atlases (\autocite{lewis20}) and dictionary of immunological terms (\autocite{japanuniversity23}).

\bigskip

Thirdly, it is strongly deterred to incorporate cell populations from different hierarchical levels in the analysis, as this may lead to increased multicollinearity or even violate the independence assumption between purified expression profiles. 
The best results are typically achieved by constructing signature matrices at the finest level of granularity, as they mitigate \enquote{dropouts} effects by better delineating closely related cell types. 

In order to compute back the contributions of the parental and higher-ranked cell lines, \autocite{sturm_etal19} provides the R function \href{https://omnideconv.org/immunedeconv/reference/map_result_to_celltypes.html}{map\_result\_to\_celltypes} in the \texttt{immunedeconv} package, which automatically aggregates estimated descendant ratios to compute the parental fraction (or even cell lines separated by further layers of lineage).

\bigskip
 Ultimately, bad characterisation of cell populations may stem from existing intra-variability within a cell population, which results from asynchronous dynamics, such as the coexistence of different phases of the cell cycle. 
 
 While in controlled conditions, such as cell cultures, chemical arrest or nutrient starvation can achieve synchronisation of the cell cycles \autocite{bar-joseph_etal08}, it becomes a challenging task when profiling living tissue \footnote{For instance, the CD3 marker, commonly used to define T cell subsets, may exhibit variable expression levels or even be entirely absent, depending on the cell cycle phase.}.
 
 Sample-specific events, such as \emph{heterotypic} contamination (for instance, infiltrates
of blood circulating immune cells, \autocite{chang_etal19}), disease-induced (\autocite{gaujoux13}) or microenvironment dysregulations (\autocite{tang_etal20}) may additionally alter the transcriptomic profiles of purified cell lines. 

\bigskip
Accordingly, to mitigate the significant loss of performance commonly observed between artificial benchmarks and real-world conditions, it is recommended to collect purified profiles in a variety of tissues, or at least representative of the phenotype condition of the bulk profiles to deconvolve \footnote{Unfortunately, this recommendation is rarely observed, for instance, the expression profile of eosinophils, in the LM22 signature of Cibersort (\autocite{newman_etal15}) was solely estimated from three distinct samples, from the same cohort.}. 
The performance of deconvolution algorithms in real conditions depends more on the representativeness of cell types profiled in the signature and environmental conditions than the choice of the regression or probabilistic framework, as discrepancies between the phenotype and tropic conditions of purified samples, compared to bulk profiles, can introduce significant bias and reduce model accuracy (\autocite{sturm_etal20}, \autocite{cai_etal22}).

\begin{figure}
\begin{subfigure}[p]{0.9\textwidth}
     \centering
         \includegraphics[width=\textwidth]{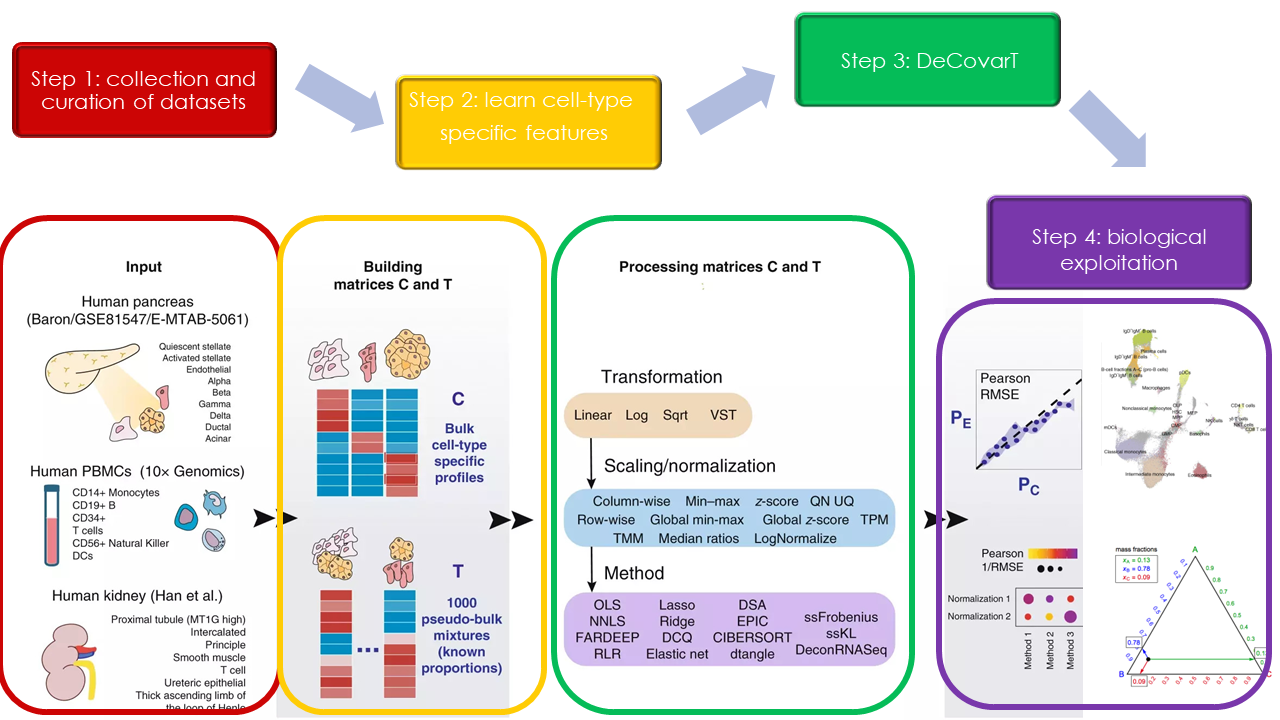}
         \label{subfig:deconvolution-pipeline}
         \caption{\textbf{Workflow for bulk deconvolution methods.}  Inspired from \autocite[Fig. 1]{avilacobos_etal18}.} 
     \end{subfigure}   
     \vfill
     \begin{subfigure}[p]{0.8\textwidth}
     \centering
         \includegraphics[width=\textwidth]{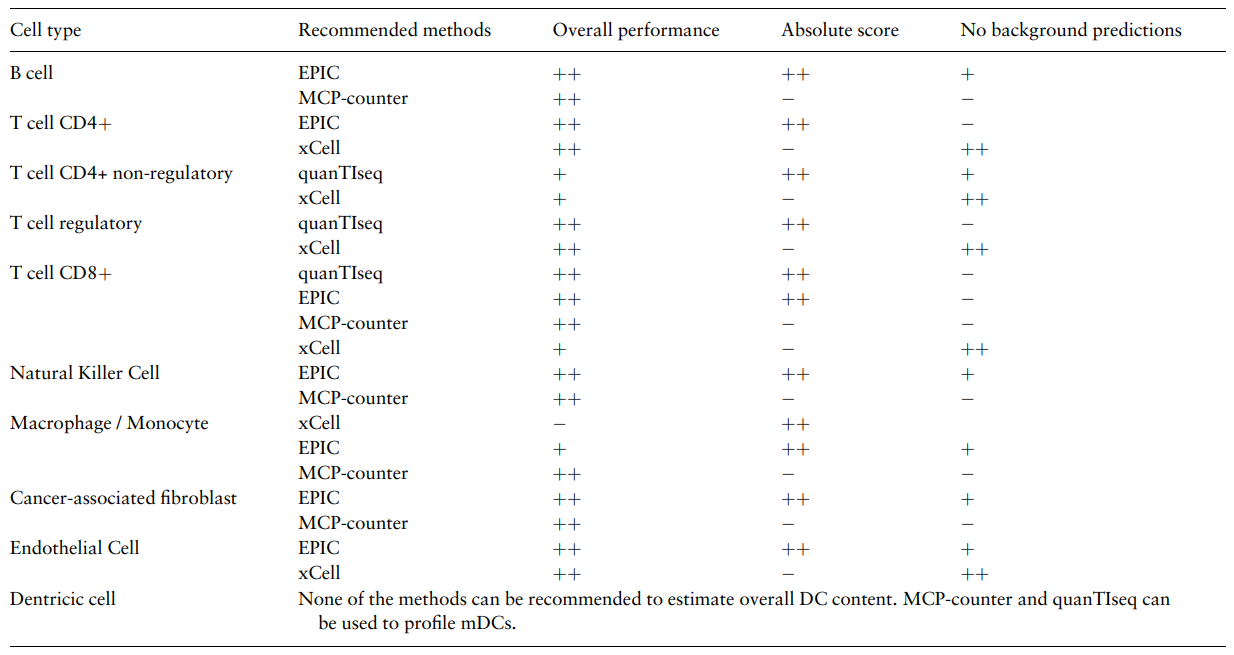}
         \label{subfig:signatures-vs-deconvolution}
         \caption{\textbf{Guidelines for the selection of a deconvolution algorithm.}  The \textit{overall} metric quantifies the correlation between the inferred fractions with the initial parameters used in the benchmark. The \textit{background prediction} is a proxy of of the inclined of a deconvolution method to forecast the presence of a cellular classification, even when absent in the mixture. Reproduced from \autocite[Table. 2]{sturm_etal19}.} 
     \end{subfigure}   
     \caption{\textbf{Outline and general guidelines for practical application of deconvolution algorithms.}}
\end{figure}

\bigskip
\bigskip
As a final note, we quote \autocite{sturm_etal19}, who believed that the \enquote{improvements made to signature matrices largely outweigh potential algorithmic improvement}. We refer the reader to \Cref{subfig:signatures-vs-deconvolution} providing general guidelines on the best signature to harness, with respect to the cell populations profiled.


On the contrary, \autocite{avilacobos_etal18} and \autocite{fa_etal20} single-cell-based deconvolution methods, capitalising on virtually reconstructed signature profiles from scRNA-Seq data, do not show significant improvement over more classical methods based on bulk-deconvolution methods.

On average, \autocite{jin_liu21} shows that penalized regression approaches, including Lasso, Ridge and Elastic Net approaches, the latter formally implemented in the DCQ algorithm \autocite{altboum_etal14}, underperformed, while on the contrary, standard OLS, see \Cref{par:linear-modelling-cellular-ratios}, and robust regression approaches (RLR, FARDEEP, SVR, see \Cref{par:feature-selection-deconvolution}) partial deconvolution methods, exhibit overall the best performances.

\bigskip
Interesting review papers encompass the works by \autocite{finotello_etal19a}, \autocite{petitprez_etal18}, \autocite{avilacobos_etal18} and \autocite{blasco_etal21}.

\section*{Perspectives: the Fate of Deconvolution Algorithms with the Development of Spatial Transcriptomics and single cell RNA-Seq}
\label{sec:general-deconvolution-perspectives}
\subsection{Overview of Spatial Transcriptomics and Single-Cell RNA Sequencing}

Spatial transcriptomics enables the simultaneous profiling of gene expression at a high spatial resolution \textit{in-situ}, while preserving the global cellular layout. ST reveals notably useful to determine the general layout of cell populations within a tissue and to identify hotspots, also known as \enquote{niches} (localised microenvironments in which stem cells prevail over fully differentiated cell subtypes) \footnote{It is common to use the abbreviation \enquote{SRT}, for Spatially Resolved Transcriptomics, when referring to the general spatial sequencing framework, in order to mitigate nomenclature confusion with the specific and corporate technology \enquote{Spatial Transcriptomics} (\autocite{stahl_etal16})/}. 

However, the design of the lattice of spots in ST technologies, such as HDST \autocite{vickovic_etal19} or Slide-Seq \autocite{rodriques_etal19}), is constrained by physical limitations that directly alleviate the final \emph{resolution} (namely the distance between capture spots). Hence, it is not uncommon that the mRNA collected at a given sport constitutes a mixture of cell types, rather than representing a single cell.

Thus, SRT techniques have to meet a middle ground between cellular resolution and the depth and coverage of the RNA library. For instance, approaches like SeqFISH+ (\autocite{eng_etal19}) and MERFISH (\autocite{chen_etal15}) provide subcellular resolution but are limited in throughput. Conversely, Spatial Transcriptomics (\autocite{stahl_etal16}) and FISSEQ (\autocite{lee_etal14}) exhibit larger coverage of the genome, yet they cannot achieve single-cell resolution sequencing and are further constrained by high detection thresholds \footnote{A minimal number of 200 mRNA molecules per cell is required to detect the expression of a transcript, excluding practically a large amount of genes involved only in specific phases of the cell cycle}. 

\bigskip
Single-cell RNA sequencing (scRNA-Seq) provides a high-resolution view of the transcriptome, by quantifying RNA content at the single-cell level. scRNA-Seq enabled to uncover cellular heterogeneity, identify rare cell populations, and capture complex dynamic changes in gene expression, that were typically obscured in bulk RNA-Seq analysis.

However, scRNA-Seq is costly and time-consuming, making it challenging to scale up for large sample sizes. In addition, the sparse nature of scRNA-Seq outputs, resulting from \enquote{drop-outs} and the complexity of the technology, renders the analysis challenging and prone to higher technical biases and variability. Hence, going down to the single cell level, scRNA-Seq typically exhibits lower coverage and depth compared to bulk RNASeq (but still higher compared to SRT).

\bigskip
Coupling scRNA-Seq with spatial transcriptomic data streamlines the understanding of the mechanisms relating gene expression patterns with changes of cell populations within tissues, by bridging the advantages of both methodologies while mitigating their major limitations.  However, \emph{mismatch}, designing the discordance between the cell types inferred from expression profiles derived from single-cell RNA sequencing and SRT, is commonly observed. Mismatch usually results from pre-sequencing and post-sequencing artefacts. Pre-sequencing mismatch can stem from \textit{sampling bias} of the tissue section (lower depth with spatial barcoding or lower access to intertwined tissue structures with HPRI) or from an artificial and ectopic stimuli perturbing the cellular expression profile (stress response, or less likely, alteration of cell phenotype due to the disruption of \textit{in situ} spatial dynamics resulting from tissue dissociation). 

\subsection{Construction of reference signatures, based on single Cell RNA-Seq profiles}
 \label{subsec:sc-DeCovarT}
On the other hand, single-cell RNA sequencing technologies empower cellular deconvolution algorithms, by enabling the derivation of signature matrices more representative of the phenotype condition. 

Indeed, by capturing gene expression profiles at the single-cell level, scRNA-Seq allows better discrimination of closely related cell types, and identification of rare cell type variants, which are likely to be confused with noise using bulk RNA-Seq. 

Even better, the stronger granularity of scRNA-Seq outputs enables to capture the heterogeneity within cell populations, including unravelling asynchronous states of a cell population. 

\subsection{Integrating Spatial Transcriptomics with Single-Cell RNA-Seq Data Through Deconvolution Approaches}
\label{subsec:sc-and-st}

 Recent alternative to mitigate the low detection threshold of scRNA in SRT and better handle mismatch issues, involve two primary approaches: \emph{deconvolution} algorithms and \emph{mapping} (report to Appendix E).  

 \bigskip
 Spatial deconvolution tools, a close synonym to \emph{stochastic profiling} techniques, estimate the cell composition for each capture spot. While sequencing the transcriptome at the single cell level is usually infeasible in a spatial context, aggregating the expression of a random pool of cells (usually rather small, aggregating no more than a dozen of them) automatically increases the depth and coverage of the RNA library, which in turn counterbalances the intrinsic noisiness and low resolution of scRNA-Seq methods. 
 
 Spatial deconvolution algorithms usually capitalise on reference signatures obtained from single-cell RNA sequencing profiles (see \cref{subsec:sc-DeCovarT}), instead of bulk expression. The final signature is finally computed by summing the individual cellular contributions in order to reconstitute a \enquote{pseudo-bulk} mixture.
 
 Nonetheless, spatial deconvolution algorithms necessitate specific adjustments compared to traditional approaches, as conventional deconvolution algorithms, designed for bulk transcriptome, often yield suboptimal results when dealing with sparse expression matrices, inherent to the SRT framework (\autocite{kleshchevnikov_etal20}). In addition, spatial deconvolution methods face similar challenges to traditional deconvolution algorithms, as they too, cannot obtain absolute estimation of cell ratios, thus limiting their applicability for meaningful intra-sample comparisons.

\begin{figure}[H]
         \centering
         \includegraphics[width=0.7\textwidth]{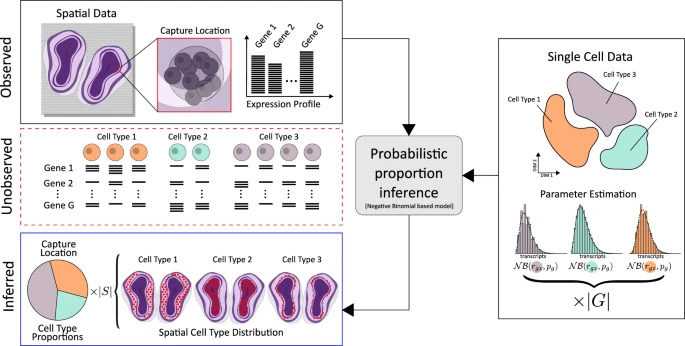}
         \label{fig:sc-deconvolution}
         \caption{\textbf{Illustration of a spatial deconvolution algorithm principle, with \texttt{stereoscope}.}A deconvolution algorithm is used to model and infer the mixture composition of cell populations at a specific capture site using signatures derived from single-cell datasets. \texttt{stereoscope} precisely employs a convolution of Negative Binomials to model the mixture of cell types within a captured side. Reproduced from \autocite[Fig. 1]{khozoie_etal21}.} 
\end{figure}

 \bigskip
 The most population spatial deconvolution methods encompass, ranked by analytical complexity:
\begin{itemize}

\item The most basic methods calculate \enquote{enrichment scores} that indicate the degree of association between an individual spatial location and a specific cell type. These scores are computed using the same techniques outlined in \Cref{subsec:marker-based-approaches}. For example, in Seurat, by \autocite{kiselev_etal17}, each spatial location is assigned to the cell type whose expression profile, composed of the markers within its gene set, exhibits the highest similarity. 

Taking a more advanced approach, the Multimodal Intersection Analysis (MIA, \autocite{moncada_etal20}) combines gene pathway information inferred from scRNA-Seq data with gene modules that are identified as enriched through spatial barcoding techniques.
	
    \item SPOTlight \autocite{elosua-bayes_etal21} and SpatialDWLS \autocite{dong_yuan21} are both regression-based models that used linear solvers to estimate cellular ratios while enforcing the unit-simplex constraint, through the non-negative least squares (NNLS) algorithm.
    
    \item \textit{Probabilistic models}, represent the mixture as a convolution of parametric distributions whose estimated cell ratios are the MLE (alternatively the MAP whereby a prior distribution is assigned to the cell ratios) of the distribution. Stereoscope (\autocite{khozoie_etal21}, also illustrated in \cref{fig:sc-deconvolution}) and Cell2location (\autocite{kleshchevnikov_etal20}) fit the distribution with a mixture of negative binomial(NB) distributions, while Robust cell-type decomposition (RCTD, \autocite{cable_etal22}) utilises Poisson distributions. 
    
    \item NMF regression (NMFref) is an unsupervised algorithm used both by SlideSeq \autocite{xu_etal16} and SPOTLight \autocite{gulati_etal13} to infer simultaneously cellular ratios and individual expression profiles.   
       
    \item More exotic and recent methods explore alternative ways, such as DSTG \autocite{he_etal20} algorithm using  \textit{mutual nearest neighbour clustering} or deep-learning methods, with Tangram \autocite{bergenstrahle_etal20}.  
\end{itemize}

Promising studies extend the investigational capacity of spatial transcriptomics, by coupling high-resolution \textit{tissue images} with \textit{histological annotations} (cell sizes and shapes, for instance) and \textit{SRT} data (\autocite{larsson_etal22}).  It is hence believed that the integration of distinct biological modalities in a spatially resolved context is poised to elucidate the as-yet-unsolved biological processes driving the spatial organisation of tissue niches(\autocite{rozenblatt-rosen_etal17}).

\texttt{SpaDecon}, by \autocite{coleman_etal23}, is one of the most promising spatial integrated approach, coupling histological annotations with metabolic and transcriptomic activity. \texttt{34P}, by \autocite{occelli_etal23}, even claims to be able to dissect intra-tumour heterogeneity in luminal breast cancer by integrating morphological annotations, SRT data and whole slide images to a neural network architecture. As a complementary resource, we refer the interested readers to \autocite{rao_etal21}, \autocite{longo_etal21}, \autocite{kresimirlukic21} and \autocite{williams_etal22} for a comprehensive and updated review of a whole array of pioneering methods integrating spatial transcriptomic, scRNA-Seq technologies and imagery annotations. 

\bigskip

To close this discussion, \autocite{teschendorff_etal17} pinpointed that the lack of reproducibility and robustness observed for most deconvolution methods may be mitigated by coupling cellular estimates obtained from distinct biological sources. Yet, a comprehensive benchmark comparing the performance of deconvolution approaches with regard to the biological input still lacks.


\clearpage
\printbibliography 
\end{document}